\documentclass[10pt,a4paper]{article}
\pdfoutput=1
\usepackage{jcappub}
\usepackage{amsmath,amssymb,graphicx,xspace,subfigure,tikz}
\usepackage{bm}
\usepackage{mathrsfs}
\usepackage{ulem}
\usepackage{cancel}
\usepackage{tensor}
\usepackage{hyperref} 
\hypersetup{
    colorlinks=true,       % false: boxed links; true: colored links
    linkcolor=red,          % color of internal links
    citecolor=blue,        % color of links to bibliography
    filecolor=magenta,      % color of file links
    urlcolor=blue           % color of external links
}
\usepackage[all]{hypcap} 

\definecolor{brown(web)}{rgb}{0.65, 0.16, 0.16}
\newcommand{\para}[1]{\textsf{\color{brown(web)}{#1}}}
\newcommand{\ba}[1]{\begin{align} #1 \end{align}}
\newcommand{\bes}[1]{\begin{equation}\begin{split} #1 \end{split}\end{equation}}
\newcommand{\bsa}[2]{\begin{subequations}\label{#1}\begin{align} #2 \end{align}\end{subequations}}
\newcommand{\com}{\;,}
\newcommand{\per}{\;.}

\newcommand{\fref}[1]{figure~\ref{#1}}
\newcommand{\Fref}[1]{Figure~\ref{#1}}

\newcommand{\sref}[1]{section~\ref{#1}}

\newcommand{\pref}[1]{(\ref{#1})}
\newcommand{\eref}[1]{eq.~(\ref{#1})}

\newcommand{\erefs}[2]{eqs.~(\ref{#1})~and~(\ref{#2})}

\newcommand{\rref}[1]{ref.~\cite{#1}}

\newcommand{\rrefs}[1]{refs.~\cite{#1}}

\newcommand{\etal}[1]{et al.}
\newcommand{\eg}[1]{e.g.}
\newcommand{\ie}[1]{i.e.}

\newcommand{\Mpl}{M_\mathrm{pl}}

\newcommand{\dd}{\mathrm{d}}
\newcommand{\ee}{\mathrm{e}}
\newcommand{\ii}{\mathrm{i}}
\newcommand{\xvec}{{\bm x}}
\newcommand{\yvec}{{\bm y}}
\newcommand{\kvec}{{\bm k}}
\newcommand{\qvec}{{\bm q}}
\newcommand{\pvec}{{\bm p}}
\newcommand{\vvec}{{\bm v}}
\newcommand{\lvec}{{\bm l}}
\newcommand{\dvec}{{\bm \nabla}}

\newcommand{\Nbb}{\mathbb{N}}
\newcommand{\Ncal}{\mathcal{N}}
\newcommand{\lfs}{l_\mathrm{fs}}
\newcommand{\kfs}{k_\mathrm{fs}}
\newcommand{\kfsast}{k_{\mathrm{fs},\ast}}
\newcommand{\qast}{q_\ast}
\newcommand{\anr}{a_\mathrm{nr}}
\newcommand{\aeq}{a_\mathrm{eq}}

\newcommand{\Scal}{\mathcal{S}}

\newcommand{\FRW}[1]{\text{$\mathrm{FRW}$}}
\newcommand{\BBN}[1]{\text{$\mathrm{BBN}$}}
\newcommand{\CMBR}[1]{\text{$\mathrm{CMBR}$}}
\newcommand{\Lya}[1]{\text{$\mathrm{Ly}\alpha$}}
\newcommand{\WWDM}[1]{\text{$\mathrm{WWDM}$}}
\newcommand{\SM}[1]{\text{\sc sm}}
\newcommand{\DM}[1]{\text{\sc dm}}
\newcommand{\DE}[1]{\text{\sc de}}
\newcommand{\X}[1]{X}
\newcommand{\LCDM}[1]{\text{$\Lambda\mathrm{CDM}$}}
\newcommand{\SE}[1]{\text{$\mathrm{SE}$}}
\newcommand{\EMD}[1]{\text{$\mathrm{EMD}$}}
\newcommand{\EDE}[1]{\text{$\mathrm{EDE}$}}
\newcommand{\vEDE}[1]{\text{$\mathrm{vEDE}$}}
\newcommand{\Gyr}{\, \mathrm{Gyr}}
\newcommand{\km}{\, \mathrm{km}}
\newcommand{\hMpc}{\, h \, \mathrm{Mpc}^{-1}}
\newcommand{\Mpc}{\, \mathrm{Mpc}}
\newcommand{\meV}{\, \mathrm{meV}}
\newcommand{\eV}{\, \mathrm{eV}}
\newcommand{\keV}{\, \mathrm{keV}}
\newcommand{\MeV}{\, \mathrm{MeV}}
\newcommand{\GeV}{\, \mathrm{GeV}}

\begin{document}

%==================================
% Title, author, and abstract
%==================================
\title{\centering 
Free streaming of warm wave dark matter in modified expansion histories
}

\author[a]{~Andrew~J.~Long}
\emailAdd{al72@rice.edu }
\author[a]{\ and \ Moira Venegas}
\emailAdd{mv58@rice.edu}
\affiliation[a]{Department of Physics and Astronomy, Rice University, Houston, Texas 77005, U.S.A.}

\abstract{
In models of warm dark matter, there is an appreciable population of high momentum particles in the early universe, which free stream out of primordial over/under densities, thereby prohibiting the growth of structure on small length scales.  The distance that a dark matter particle travels without obstruction, known as the free streaming length, depends on the particle's mass and momentum, but also on the cosmological expansion rate.  In this way, measurements of the linear matter power spectrum serve to probe warm dark matter as well as the cosmological expansion history.  In this work, we focus on ultra-light warm wave dark matter (WWDM) characterized by a typical comoving momentum $q_\ast$ and mass $m$.  We first derive constraints on the WWDM parameter space $(q_\ast, m)$ using Lyman-$\alpha$ forest observations due to a combination of the free-streaming effect and the white-noise effect.  We next assess how the free streaming of WWDM is affected by three modified expansion histories: early matter domination, early dark energy, and very early dark energy.  
}

\keywords{
warm dark matter, free streaming, matter power spectrum
}

\maketitle

%Add an empty line between paragraphs
%\setlength{\parindent}{20pt}
\setlength{\parskip}{2.5ex}

%==================================
% Introduction
%==================================
\section{Introduction}
\label{sec:intro}

%=========
Often when we think about ultra-light dark matter \cite{Antypas:2022asj}, we imagine that it's extremely cold.  
``Coldness'' can be quantified as the dark matter's velocity dispersion at the time of production, $\langle |\vvec|^2 \rangle = \langle |\pvec|^2 / E(\pvec)^2 \rangle$, as compared with $c^2 = 1$.  
Equating ``ultra-light'' and ``ultra-cold'' is the correct intuition for some of the simplest scenarios of dark matter production.  
Misalignment production, which may be the origin of axion dark matter~\cite{Preskill:1982cy,Abbott:1982af,Dine:1982ah}, arguably leads to the coldest possible dark matter since nonzero-momentum modes are only populated by the field's tiny quantum fluctuations during inflation~\cite{Kolb:2023ydq}.  
However, there a great variety of theoretically-compelling and phenomenologically-rich scenarios \cite{Graham:2015rva,Gorghetto:2018myk,Agrawal:2018vin,Dror:2018pdh,Co:2018lka,Bastero-Gil:2018uel,Long:2019lwl,Gorghetto:2020qws,Co:2021rhi,Redi:2022llj,Harigaya:2022pjd} in which ultra-light dark matter is not produced ultra-cold.  
For instance, if the dark matter is produced after inflation from the decay of a metastable cosmological relic, like a moduli field or a topological defect network, then it may even be ultra-relativistic at the time of production.  

%=========
These models in which the dark matter is ultra-light but not ultra-cold~\cite{Amin:2022nlh} have attracted attention recently, where they have been called ``warm and fuzzy dark matter''~\cite{Liu:2024pjg} and ``warm wave dark matter'' (\WWDM{})~\cite{Ling:2024qfv}. 
If the dark matter's mass is around $m = 10^{-20} \eV$ and the dark matter's characteristic comoving momentum is around $\qast = 10^3 \hMpc \sim 10^{-6} m$, then the models are expected to leave their imprint on the small-scale spatial distribution of matter.  
In this regard, the phenomenology of \WWDM{} is distinguished from cold dark matter by two effects: the white-noise effect and the free-streaming effect.  
On the one hand, the high-momentum population of \WWDM{} particles corresponds to small-scale inhomogeneities in the dark matter density field, which can be understood as a white-noise component in the dark matter power spectrum \cite{Amin:2022nlh}. 
On the other hand, if the dark matter's initial velocity dispersion is sufficiently large, then it will free stream out of over-densities and suppress the growth of cosmological structures on small length scales.  
Depending on the parameters, the free streaming of warm wave dark matter will lead to a suppression of the dark matter inhomogeneities on length scales ($k \approx \mathrm{few} \hMpc$) that are probed by current and future surveys.  
Such observations have already provided powerful constraints~\cite{Boyarsky:2008xj,Viel:2013fqw,Villasenor:2022aiy,Arias:2023wyg} on thermal relic warm dark matter, such as sterile neutrinos with a mass at the keV-scale.  

\begin{figure}[t]
\centering
\includegraphics[width=0.75\textwidth]{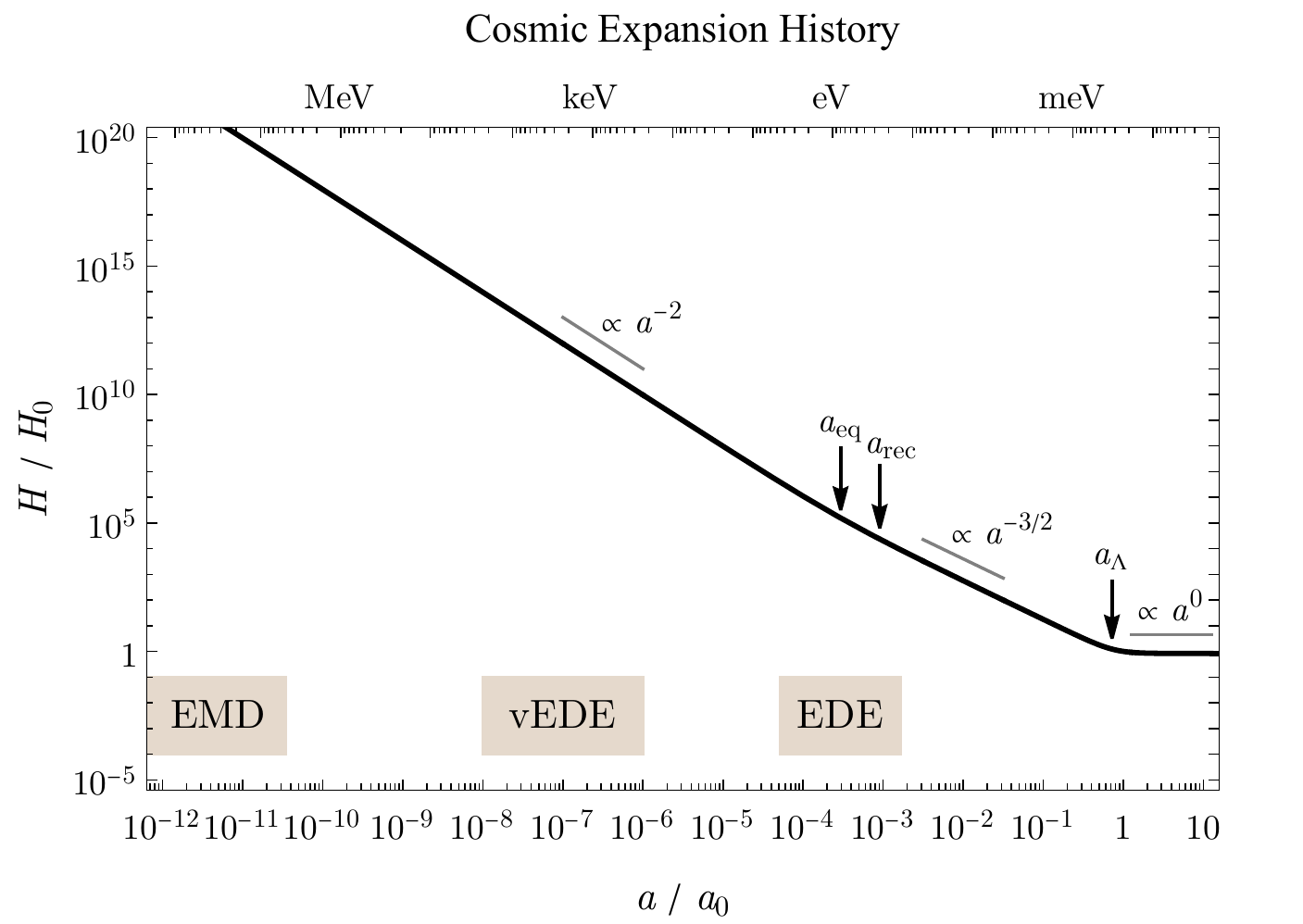}
\caption{\label{fig:timeline}
An illustration of the late-time cosmological expansion history and possible modifications. We show the cosmological expansion rate $H(t)$ (normalized to its value today $H_0$) as a function of the scale factor $a(t)$ (normalized to its value today $a_0$) over a period of time $t$ during which the expansion history might be modified.  In \LCDM{} Cosmology the expansion rate scales as $a^{-2}$ during Standard Model radiation domination, as $a^{-3/2}$ during dark matter domination, and as $a^{0}$ during dark energy domination.  The top axis shows the corresponding photon temperature $T(t) \propto a(t)^{-1}$. In this work we consider several alternative expansion histories that modify the relationship between $H$ and $a$ at different times: early matter domination (\EMD{}), very early dark energy (\vEDE{}), and early dark energy (\EDE{}).  
}
\end{figure}

%=========
In this work we explore the idea that the free streaming of warm wave dark matter affords a powerful probe of the expansion history of the universe.  
In the cosmological concordance model, \ie{} \LCDM{} Cosmology, the cosmological energy budget is shared among Standard Model particles and radiation, dark matter, and dark energy~\cite{Baumann:2022mni}.  
Together these various energy sources govern the cosmological expansion rate $H(t)$ through Einstein's equation (in the form of Friedmann's equations).  
Models in which $H(t)$ differs from the \LCDM{} prediction are collectively known as modified expansion histories~\cite{Allahverdi:2020bys,Batell:2024dsi}.
A few examples are illustrated in \fref{fig:timeline}.  
For the most part, \LCDM{} Cosmology works extremely well to explain precision cosmological observations such as the cosmic microwave background radiation (\CMBR{}) anisotropies and large scale structure. 
However, several anomalies suggest that a modified expansion history may be needed~\cite{Perivolaropoulos:2021jda}.  
Some of these include discrepant measurements of the present-day cosmological expansion rate (\ie{} the Hubble tension)~\cite{DiValentino:2021izs,Freedman:2021ahq} and evidence for evolving dark energy inferred from observations made by the Dark Energy Spectroscopic Instrument (DESI) and others~\cite{DESI:2024mwx}.  
Moreover, modified expansion histories are a generic expectation for theories with new physics, since heavy metastable particles (like string theory moduli) might come to dominate the energy budget before decaying \cite{Dienes:2011ja,Kane:2015jia,Dienes:2021woi,Cicoli:2023opf}. 
Therefore, it's important to develop probes of the expansion history using all available tools.  

%=========
In this work we investigate how modified expansion histories impact the free streaming of warm wave dark matter, since the distance $\lfs$ traveled while free streaming is calculated as a time integral over the cosmic history.  
Our perspective is that a future detection of dark matter in laboratories on Earth will furnish measurements of $m$ and $q_\ast$, perhaps because the signal is proportional to the squared field amplitude that oscillates with angular frequency $\omega \approx m$ and maintains phase coherence on a time scale $t_c \approx 1/m\qast^2$ \cite{Graham:2015ifn}.  
Supposing that these measurements allow $\lfs$ to be predicted accurately in the standard cosmology, then any mismatch between the predicted free streaming scale and observations could be interpreted as evidence of a modified expansion history.\footnote{With a similar motivation to ours, the authors of \rrefs{Gelmini:2008sh,Cohen:2008nb,Chung:2011hv,Chung:2011it} study how a mismatch in the predicted and observed properties of thermal relic dark matter could probe the early universe cosmic expansion history.}

We assess the extent to which measurements of the linear matter power spectrum at small length scales can probe this scenario. 

%=========
The remainder of this article is organized as follows.  
In \sref{sec:WWDM} we review how warm wave dark matter may be described by a power spectrum for the field amplitude and how the linear matter power spectrum is calculated, including the white-noise efect and the free-streaming effect. 
In \sref{sec:model_0} we present constraints on the \WWDM{} parameter space using measurements of the linear matter power spectrum derived from \Lya{} observations assuming a standard cosmological expansion history.  
Next we explore the free streaming of \WWDM{} in a few examples of modified cosmological expansion histories: early matter domination in \sref{sec:model_1}, early dark energy in \sref{sec:model_2}, and very early dark energy in \sref{sec:model_3}.  
Finally we summarize and conclude in \sref{sec:conclusion}. 

%=================
% --- FIGURE 1 --- 
%=================

%==================================
% Warm wave dark matter
%==================================
\section{Warm wave dark matter}
\label{sec:WWDM}

%=========
\para{Cosmological expansion.}  
The background (unperturbed) spacetime is homogeneous and isotropic with vanishing spatial curvature.  
The Friedmann-Robertson-Walker (\FRW{}) metric's spacetime interval is 
\ba{\label{eq:FRW_spacetime}
    (\dd s)^2 = (\dd t)^2 - a(t)^2 \, |\dd \xvec|^2 
    \qquad \text{and} \qquad 
    a(t) \, \dd \tau = \dd t 
    \com
}
where $\xvec = (x,y,z)$ are the comoving spatial coordinates, $t$ is the time coordinate, and $\tau$ is the conformal time coordinate.  
The FRW scale factor $a(t)$ increases at a rate given by the Hubble parameter $H(t) = \dd \ln a / \dd t$.  
The Hubble parameter satisfies the first Friedmann equation 
\ba{\label{eq:Friedmann_eqn}
    3 \Mpl^2 H(t)^2 = \rho_\SM{}(t) + \rho_\DM{}(t) + \rho_\DE{}(t) + \rho_\X{}(t) \equiv \rho_\mathrm{tot}(t) 
    \com
}
where $\Mpl = 1/\sqrt{8 \pi G_N} \approx 2.435 \times 10^{18} \GeV$ is the reduced Planck mass. 
The $\rho$'s are explained below. 

%=========
\para{Energy budget.}  
The cosmological expansion is induced by the various components of the cosmological energy budget.  
Their respective energy densities are denoted by $\rho_\SM{} = \rho_\mathrm{b} + \rho_\mathrm{r}$ for the Standard Model particles and radiation, $\rho_\DM{}$ for the dark matter, $\rho_\DE{}$ for the dark energy (assumed to be a cosmological constant), and $\rho_\X{}$ for a new energy component that will induce a period of modified cosmic expansion.  
After the epoch of electron-positron annihilations, when the plasma temperature is $T \lesssim 0.05 \MeV$, the effective number of relativistic species does not change significantly~\cite{Husdal:2016haj}.  
Assuming that the dark matter is already nonrelativistic by this time, the energy densities are approximately 
\bsa{eq:rho_SM_DM_Lambda}{
    \rho_\SM{}(t) & = \Omega_\mathrm{b} \, \rho_\mathrm{crit} \, \bigl( a(t) / a_0 \bigr)^{-3} + \Omega_\mathrm{r} \, \rho_\mathrm{crit} \, \bigl( a(t) / a_0 \bigr)^{-4} \\ 
    \rho_\DM{}(t) & = \Omega_\mathrm{c} \, \rho_\mathrm{crit} \, \bigl( a(t) / a_0 \bigr)^{-3} \\ 
    \rho_\DE{}(t) & = \Omega_\Lambda \, \rho_\mathrm{crit} 
    \com
}
where $a_0 \equiv a(t_0)$ is the value of the scale factor today ($t = t_0$), where $\rho_\mathrm{crit} \equiv 3 \Mpl^2 H_0^2$ is the cosmological critical energy density today, where $H_0 = h H_{100}$ is the Hubble constant, and where $H_{100} = 100 \km / \mathrm{sec} / \mathrm{Mpc}$.  
The $\Omega_\mathrm{b}$ term accounts for the baryonic matter (electrons, protons, and nuclei); the $\Omega_\mathrm{r}$ term accounts for the radiation (photons and neutrinos); the $\Omega_\mathrm{c}$ term accounts for the dark matter; and the $\Omega_\Lambda$ term accounts for the cosmological constant.  
The parameters of the \LCDM{} Cosmology~\cite{Fixsen:1996nj,Planck:2018vyg} are measured to be 
%\AL{See the right-most column of Tab 2 of \rref{Planck:2018vyg}}
$T_0 = 0.2351 \pm 0.0003 \meV$, 
$\Omega_\mathrm{b} h^2 = 0.02242 \pm 0.00014$, 
$\Omega_\mathrm{r} h^2 = (4.171 \pm 0.02) \times 10^{-5}$, 
$\Omega_\mathrm{c} h^2 = 0.11933 \pm 0.00091$, 
$h = 0.6766 \pm 0.0042$, 
$\Omega_\Lambda = 0.6889 \pm 0.0056$, 
%$\Omega_\mathrm{m} = 0.3111 \pm 0.0056$, 
%$\Omega_\mathrm{m} h^2 = 0.14240 \pm 0.00087$, 
$t_0 = 13.787 \pm 0.020 \Gyr$, 
%$z_\mathrm{eq} = 3387 \pm 21$, 
%$k_\mathrm{eq} = 0.010339 \pm 0.000063 \Mpc^{-1}$, 
and we are free to take $a_0 = 1$.
%and $g_{\ast S,0} = 3.91$.  
We make use of \eref{eq:rho_SM_DM_Lambda} for most of the analysis, except for the model in \sref{sec:model_1} where we are interested in temperatures above $0.05 \MeV$.  

\begin{figure}[t]
\centering
\includegraphics[width=0.75\textwidth]{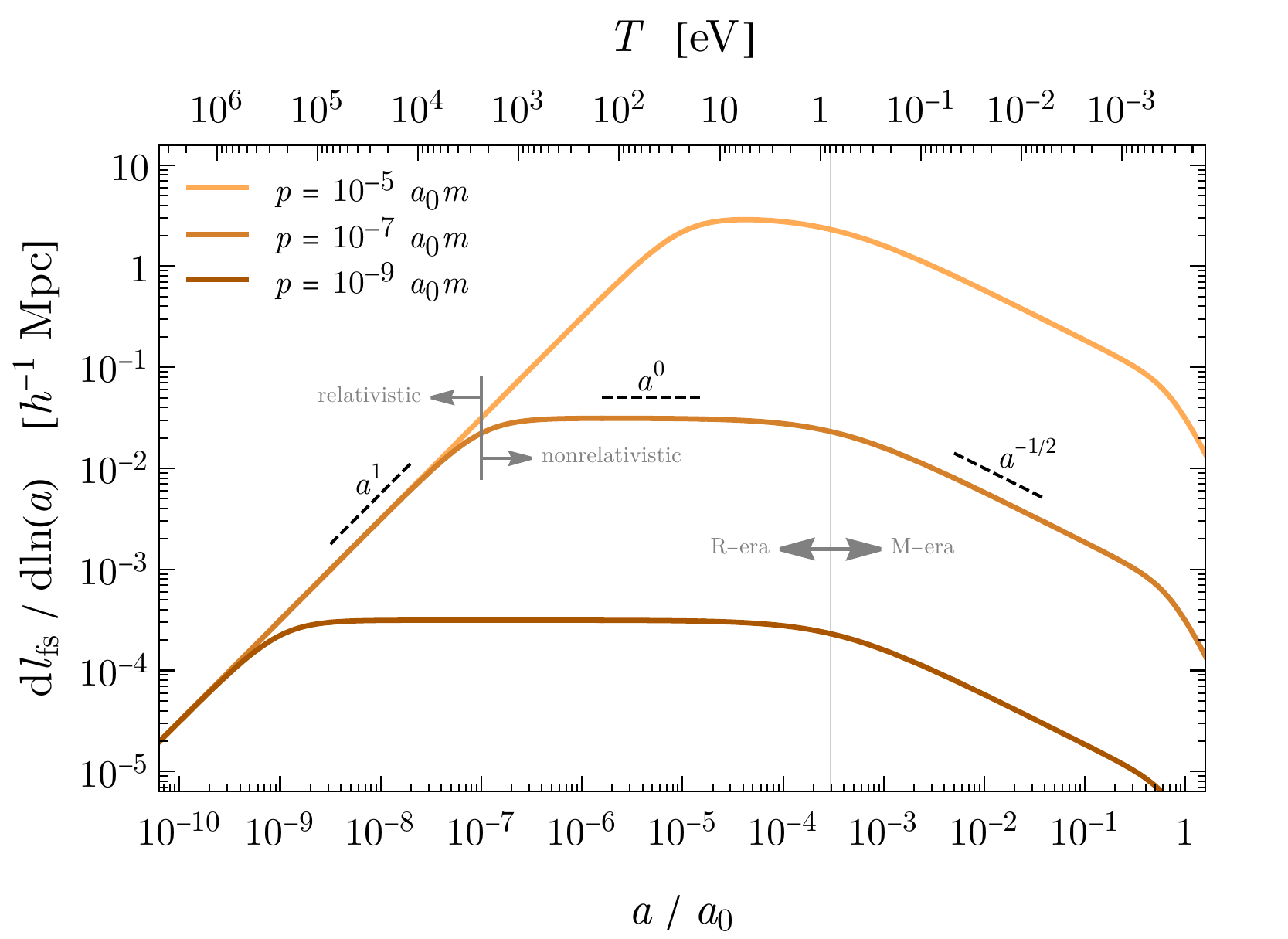}
\caption{\label{fig:dlfs_dlna}
Contributions to the free streaming length integral.  We show the integrand $\dd l_\mathrm{fs} / \dd \ln(a)$ as a function of scale factor $a$.  The three curves correspond to different values of the comoving momentum $p$ in units of the particle mass $m$.  We highlight the divide between the radiation era (R-era) and matter era (M-era) at $z_\mathrm{eq} = 3387$.  For the middle curve ($p = 10^{-7} a_0 m$) we indicate the divide between relativistic ($p > am$) and nonrelativistic ($p < am$) particles.  We also illustrate how $\dd l_\mathrm{fs} / \dd ln(a) \propto a^1$ for relativistic particles in the radiation era, $\propto a^0$ for nonrelativistic particles in the radiation era, and $\propto a^{-1/2}$ for nonrelativistic particles in the matter era.  The free streaming length mainly accumulates while particles are nonrelativistic in the radiation era, corresponding to the horizontal branches of the plot.  
}
\end{figure}

%=========
\para{Free streaming.}  
We are interested in the free streaming of dark matter particles.  
A particle of mass $m$ with momentum $\pvec_\mathrm{phys}(t)$ and energy $E(t) = [\pvec_\mathrm{phys}(t)^2 + m^2]^{1/2}$ at time $t$ is traveling with velocity $\vvec(t) = \pvec_\mathrm{phys}(t) / E(t)$.  
If the particle is free, then its comoving momentum $\pvec = a(t) \pvec_\mathrm{phys}(t)$ is constant, and the velocity can also be written as $\vvec(t) = \pvec / [\pvec^2 + a(t)^2 m^2]^{1/2}$.  
We define the comoving free streaming displacement, denoted by $\lvec_\mathrm{fs}(\pvec,t)$, and the comoving free streaming length, denoted by $l_\mathrm{fs}(p,t)$.  
The comoving distance that a free particle with comoving momentum $\pvec$ travels between conformal time $\tau = 0$ and conformal time $\tau$ is calculated by integrating the velocity:\footnote{In principle the lower limit of integration should be chosen to be the time at which dark matter production occurs, but generally the integral is dominated by the late-time part of the integration domain, and the error that's introduced by setting the initial time to zero is negligible.} 
\bes{\label{eq:lfs_def}
    \lvec_\mathrm{fs} 
    = \int_0^{\tau} \! \dd \tau^\prime \, \vvec(t(\tau^\prime)) 
    = \int_0^{t} \! \dd t^\prime \, \frac{1}{a(t^\prime)} \ \frac{\pvec}{\sqrt{\pvec^2 + a(t^\prime)^2 m^2}} 
    = \int_0^{a} \! \frac{\dd a^\prime}{a^\prime} \, \frac{1}{a^\prime \, H(t(a^\prime))} \ \frac{\pvec}{\sqrt{\pvec^2 + (a^\prime)^2 m^2}} 
    \com
}
and $l_\mathrm{fs} \equiv |\lvec_\mathrm{fs}|$.  
In \fref{fig:dlfs_dlna} we plot the integrand assuming Standard Expansion Cosmology in order to illustrate how different periods of time contribute to the comoving free streaming length.  
For particles that become nonrelativistic in the radiation era (\ie{}, most of the dark matter), the free streaming length mainly accumulates while these particles are nonrelativistic ($a > \anr \equiv p / m$) in the radiation era ($a < \aeq$).  
Consequently the comoving free streaming length (assuming Standard Expansion Cosmology) is approximately 
\bes{\label{eq:lfs_t0}
    l_\mathrm{fs}(t_0) \approx  (p / \aeq m) \, (\aeq H_\mathrm{eq})^{-1} \, \log(2 \aeq / \anr) 
    \qquad \text{for $p \ll \aeq m$}
    \com
}
where $(\aeq H_\mathrm{eq})^{-1} \approx 68.3 \, h^{-1} \, \mathrm{Mpc}$.  
Notice that the comoving free streaming length \eqref{eq:lfs_def} depends on the cosmological expansion history through the factor of $H(t)$.  
For example, a larger expansion rate due to an additional energy component $\rho_\X{}$ in \eref{eq:Friedmann_eqn}, would reduce the free streaming length.  

%=================
% --- FIGURE 2 --- 
%=================

%=========
\para{Momentum distribution.}  
The system of interest consists of many dark matter particles with different momenta.  
We characterize this system using a one-particle phase space distribution function $F(\xvec, \pvec, t)$.  
The differential $\dd \Nbb = F(\xvec,\pvec,t) \, \dd^3 \xvec \, \dd^3 \pvec / (2\pi)^3$ gives the average number of particles with comoving spatial coordinate between $\xvec$ and $\xvec + \dd \xvec$ and with comoving momentum between $\pvec$ and $\pvec + \dd \pvec$ at time $t$. 
It is normalized such that $\int \! \dd \Nbb = N$, where $N$ is the average number of particles in the system (or in a fiducial finite volume $V$ if the system is infinite).  
If the system is statistically homogeneous, then $F(\xvec,\pvec,t)$ is independent of $\xvec$, and if it is statistically isotropic then $F(\xvec,\pvec,t)$ is independent of the orientation of $\pvec$ and only depends on $|\pvec| \equiv p$.  
Under these assumptions, the distribution function can be written as $F(\xvec,\pvec,t) = f(p,t)$. 
We can use the distribution function to calculate ensemble averages.  
For example the average energy per particle at time $t$ is evaluated as 
\bes{\label{eq:average_def}
    \langle E(t) \rangle 
    & = \int \! \frac{\dd \Nbb}{N} \ E(t)
    = \frac{1}{N} \int \! \dd^3 \xvec \int \! \! \frac{\dd^3 \pvec}{(2\pi)^3} \, F(\xvec, \pvec, t) \, E(t) 
    = \frac{V}{N} \int_0^\infty \! \frac{\dd p}{p} \frac{p^3}{2\pi^2} \, f(p, t) \, \sqrt{p^2 / a(t)^2 + m^2} 
    \com
}
where $N/V = \int \! \dd \Nbb / V = \int_0^\infty \! \frac{\dd p}{p} \frac{p^3}{2\pi^2} \, f(p, t)$ is the average number density.  

%=========
\para{Thermal relic warm dark matter.}  
The standard example of warm dark matter is thermal relic particle dark matter with mass on the order of $m \sim \mathrm{keV}$.  
If the dark matter is produced by thermal freeze out or thermal freeze in, then it inherits the momentum distribution of the plasma from which it was created.  
In other words, the phase space distribution function $f(p,t)$ is approximately a Bose-Einstein or Fermi-Dirac thermal distribution, depending on the quantum statistics (\ie{}, the spin) of the dark matter particles, and the initial temperature of the distribution is roughly the temperature of the plasma at the time of dark matter production.  
Subsequently, both the plasma and the decoupled dark matter redshift until today.  
Consequently, the majority of the dark matter particles have a comoving momentum $p$ that is comparable to the momentum of the \CMBR{} photons today, $p_\text{\CMBR{}} \approx 3 a_0 T_0 \approx 0.7 \meV$.  
For $m = 1 \keV$ the corresponding free streaming length is $l_\mathrm{fs} \approx 2 h^{-1} \Mpc$, and it approximately scales as $l_\mathrm{fs} \propto m^{-1}$.  
Since the free streaming would lead to a suppression of structure on scales below $l_\mathrm{fs}$, which is not observed in the data on scales down to $\sim \mathrm{Mpc}$, it follows that the mass of thermal relic dark matter must be larger than $\approx \mathrm{few} \times \mathrm{keV}$ \cite{Boyarsky:2008xj,Viel:2013fqw,Villasenor:2022aiy}.  

%=================
% --- FIGURE 3 --- 
%=================
\begin{figure}[t]
\centering
\includegraphics[width=0.45\textwidth]{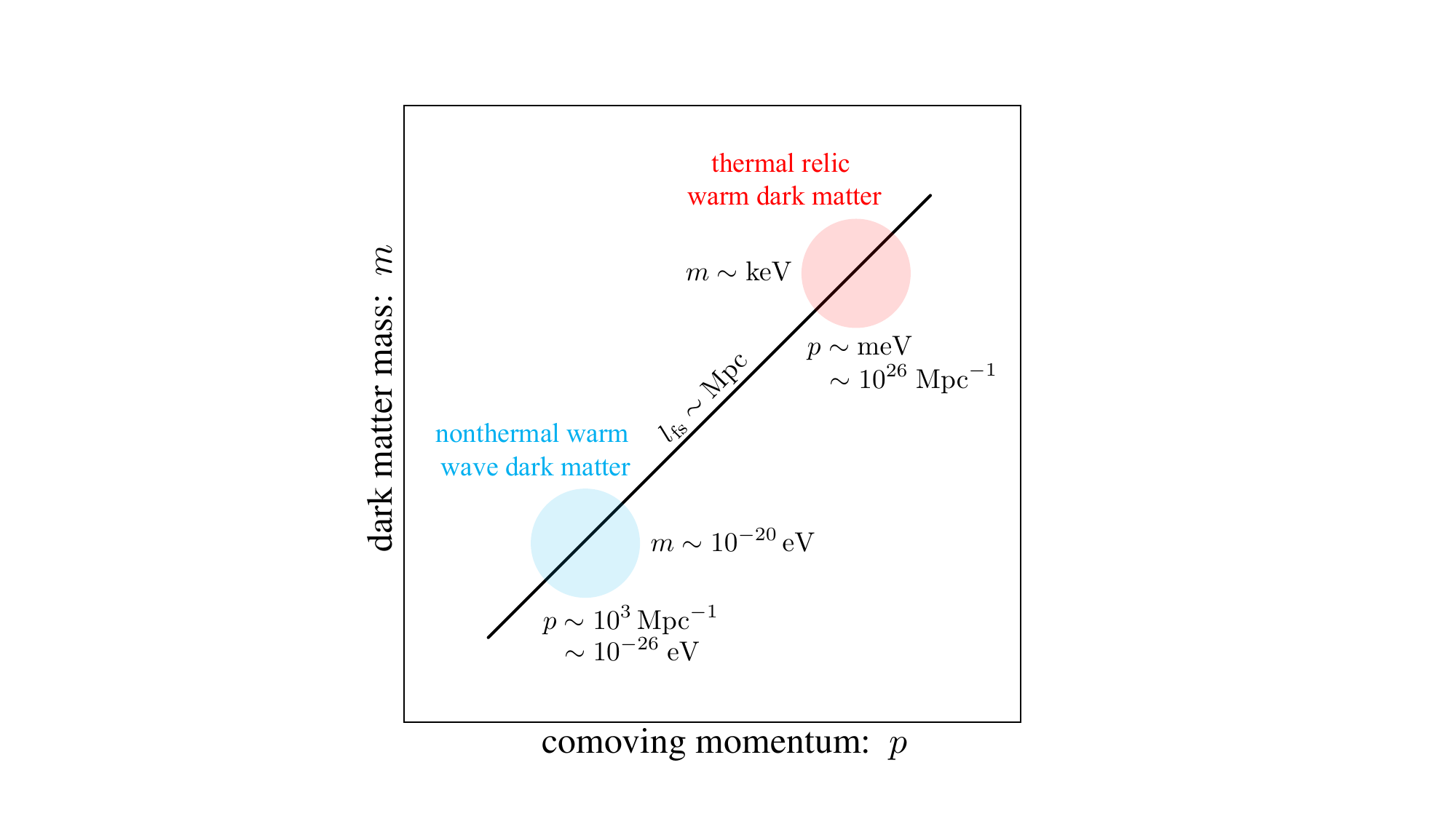}
\caption{\label{fig:WDM_parameter_space}
A global picture of the parameter space for warm dark matter.  For thermal relic warm dark matter the typical comoving momentum is around the \CMBR{} temperature today $p \sim \mathrm{meV}$, and a comoving free streaming length of $l_\mathrm{fs} \sim \mathrm{Mpc}$ is obtained for $m \sim \mathrm{keV}$.  For nonthermal warm wave dark matter with mass in the fuzzy regime $m \sim 10^{-20} \eV$, the same free-streaming length is obtained at $p \sim 10^3 \Mpc^{-1}$.  
}
\end{figure}

%=========
\para{Nonthermal warm wave dark matter.}  
If the dark matter is not produced from the plasma, but rather it is produced nonthermally so as to be colder than the plasma, then it can be much lighter than $m \approx \mathrm{keV}$ while nevertheless maintaining an acceptably small free streaming length.  
For example, fuzzy dark matter~\cite{Hu:2000ke} postulates an ultra-light dark matter candidate \cite{Hui:2016ltb} with mass around $m \sim 10^{-22} \eV$.  
A typical production mechanism for spin-0 fuzzy dark matter is the misalignment mechanism, familiar from studies of axions \cite{Preskill:1982cy,Abbott:1982af,Dine:1982ah}, which produces ultra-cold dark matter particles with negligible free streaming ($\lfs \ll \mathrm{Mpc}$).  
However, other production mechanisms create dark particles with appreciable momentum, perhaps even allowing them to be relativistic just after production. 
For example, the emission of ultra-light dark particles from a topological defect network consisting of cosmic strings, which has been studied for axions~\cite{Buschmann:2021sdq} and dark photons~\cite{Long:2019lwl}, leads to a power-law momentum distribution that extends from the scale of the string loops to the scale of the string thickness.  
For fuzzy dark matter with mass $m \sim 10^{-20} \eV$ and comoving momentum $p \sim 10^3 \hMpc$, the comoving free streaming length is $\lfs \approx \mathrm{Mpc}$, which is comparable to the typical value for thermal relic warm dark matter.  
\Fref{fig:WDM_parameter_space} illustrates the global perspective on warm dark matter, across the vast parameter space between thermal relic warm dark matter and nonthermal warm wave dark matter.  

%=========
\para{Field description.}  
For the work presented in this article, we adopt the following model of warm wave dark matter.  
We restrict our attention to spin-0 dark matter particles that arise as the quantum excitations of a real scalar field $\phi(\xvec,t)$.  
We are interested in models for which the mass of $\phi$ is on the order of $10^{-22}$ to $10^{-18} \eV$, which is known as the ``fuzzy'' or ``wave-like'' regime.  
For the systems of interest, the field has a large occupation number in the dominant Fourier modes, and it can be treated as a classical field.  
We assume that the evolution is deterministic and that the initial conditions (at time $t = t_i$) are random corresponding to stochastic fluctuations that are Gaussian with zero mean $\langle \phi(\xvec,t_i) \rangle = 0$, statistically homogeneous, and statistically isotropic.  
To study scalar inhomogeneities we work in the Newtonian gauge~\cite{Baumann:2022mni}.  
Then the statistics of the initial field configuration are fully characterized by the two-point correlation function 
\ba{\label{eq:P_phi_phi}
    \langle \phi(\xvec,t_i) \, \phi(\yvec,t_i) \rangle 
    = \int \! \! \frac{\dd^3 \qvec}{(2\pi)^3} \, P_{\phi}(|\qvec|,t_i) \, \ee^{\ii \qvec \cdot (\xvec - \yvec)} 
    = \int_0^\infty \! \frac{\dd q}{q} \, \frac{q^3}{2\pi^2} \, P_{\phi}(q,t_i) \ \mathrm{sinc}\bigl( q \, |\xvec - \yvec| \bigr) 
    \com
}
where $\mathrm{sinc}(z) = 1$ for $z= 0$ and $\mathrm{sin}(z) / z$ for $z \neq 0$.  
Unlike scalar field dark matter produced by misalignment, for which the dominant Fourier mode has $\qvec = 0$, \WWDM{} is produced with a characteristic nonzero comoving wavenumber $|\qvec| = \qast$; see \sref{sec:intro} for examples of \WWDM{} production mechanisms.  
Provided that it is sufficiently tightly peaked at $q_\ast$, the shape of the field's power spectrum does not significantly impact the linear matter power spectrum. 
This is because the free-streaming effect is controlled by the most abundant particles with $|\qvec| \approx \qast$ \cite{Liu:2024pjg}, and the white-noise effect only needs $q^3 P_\phi$ to approach zero for small $q$ (see after \eref{eq:Pdelta_from_Pphi}).  
For simplicity and concreteness, we follow \rref{Amin:2022nlh} and model the initial power spectrum $P_{\phi}(q,t_i)$ as a broken power law 
\ba{\label{eq:broken_power_law}
    \frac{q^3}{2\pi^2} P_{\phi}(q,t_i) = A_i \biggl\{ \biggl( \frac{q}{\qast} \biggr)^\nu \, \Theta(\qast - q) + \biggl( \frac{q}{\qast} \biggr)^{-\alpha} \, \Theta(q - \qast) \biggr\} 
    \com
}
where $\Theta(x) = 1$ for $x \geq 0$ and $0$ for $x < 0$.  
In order for certain integrals to be convergent (see below) the indices must be $\nu > 3/2$ and $\alpha > 2$.
We set $\nu = \alpha = 3$, and consequently the scaled power spectrum $q^3 P_{\phi}$ is tightly peaked at $q = \qast$, which implies that our results are insensitive to the precise values of these exponents.  
We are interested in values of $\qast$ above $10^2 \hMpc$.  
The prefactor $A_i$ is related to the average dark matter energy density $\bar{\rho}$ (see below), but it doesn't impact the transfer function, which is calculated as a ratio.

%=========
\para{Bridging particle and field descriptions.}  
We have discussed how the statistically-homogeneous and isotropic system can be described as a collection of randomly-distributed particles with phase space distribution function $f(p,t)$ or as a stochastic scalar field with power spectrum $P_{\phi}(q,t)$.  
In the former description, particles are labeled by a comoving momentum $\pvec$ and they carry an energy $E(p,t) = (p^2/a(t)^2 + m^2)^{1/2}$ where $p = |\pvec|$.  
In the latter description, Fourier modes of the field are labeled by a comoving wavevector $\qvec$ and they oscillate at an angular frequency $\omega(q,t) = (q^2 / a(t)^2 + m^2)^{1/2}$ where $q = |\qvec|$.  
One can bridge these two descriptions by comparing the energy and momentum densities carried by the particle population and the stochastic scalar field.  
For instance, $\dd \rho = a(t)^{-3} \tfrac{\dd^3 \pvec}{(2\pi)^3} \, E(p,t) \, f(p,t)$ gives the average energy density of particles for which the comoving momentum is between $\pvec$ and $\pvec + \dd \pvec$ at time $t$.  
Similarly, if $\rho_\phi$ is given by \eref{eq:rho_phi_def} then $\dd \rho = \tfrac{\dd^3 \qvec}{(2\pi)^3} \, \omega(q,t)^2 \, P_\phi(q,t)$ gives the average energy density of waves for which the comoving wavevector is between $\qvec$ and $\qvec + \dd \qvec$ at time $t$. 
These considerations motivate the relationship:
\ba{\label{eq:bridge}
    f(p,t) & = a(t)^3 \, \omega(q,t) \, P_{\phi}(q,t) \Bigr|_{q = p} 
    \per
}
One may be familiar with similar arguments in electromagnetism, which allow a classical electromagnetic field to be viewed as a coherent collection of photons.  

%=========
\para{Dark matter energy density.}  
We assume that the field's energy density is given by 
\bes{\label{eq:rho_phi_def}
    \rho_\phi(\xvec,t) = \tfrac{1}{2} \bigl( \partial_t \phi(\xvec,t) \bigr)^2 + \tfrac{1}{2} a(t)^{-2} \bigl| \dvec \phi(\xvec,t) \bigr|^2 + \tfrac{1}{2} m^2 \phi(\xvec,t)^2 
    \com
}
which is a reasonable expectation if non-gravitational interactions can be neglected. 
Since the stochastic field $\phi(\xvec,t)$ is a random variable, the energy density $\rho_\phi(\xvec,t)$ is also random, and on average\footnote{Note that we could have arrived at the same formula via the particle description by using the bridge relationship in \eref{eq:bridge}.  Namely, $\bar{\rho}(t) = \langle E(p,t) \rangle / V_\mathrm{phys} = a(t)^{-3} \int (\dd^3 \pvec / (2\pi)^3) E(|\pvec|,t) f(|\pvec|,t) = \int (\dd^3 \qvec / (2\pi)^3) \omega(|\qvec|,t)^2 P_{\phi}(|\qvec|,t)$.} 
\bes{
    \bar{\rho}(t)
    & \equiv \langle \rho_\phi(\xvec,t) \rangle 
    = \int \! \! \frac{\dd^3 \qvec}{(2\pi)^3} \, \omega(q,t)^2 \, P_{\phi}(|\qvec|,t) 
    = \int_0^\infty \! \frac{\dd q}{q} \, \frac{q^3}{2\pi^2} P_{\phi}(q,t) \, \omega(q,t)^2 
    \\ & 
    = (\alpha + \nu) \biggl( \frac{\qast^2 / a(t)^2}{(\alpha - 2) \, (\nu + 2)} + \frac{m^2}{\alpha \nu} \biggr) A(t) 
    \per
}
Provided that $\qast \ll \aeq m$ then the dominate Fourier modes will be nonrelativistic at the time of radiation matter equality ($t = t_\mathrm{eq}$ and $\aeq = a(t_\mathrm{eq})$).  
Then $\phi$ is a viable dark matter candidate, and we can identify $\bar{\rho}(t)$ with the homogeneous (volume-averaged) dark matter energy density $\rho_\DM{}(t)$.  

%=========
\para{Linear matter power spectrum.}  
To quantify the amplitude of dark matter inhomogeneities, we use the linear matter power spectrum $P_m(k)$.  
If the dark matter's energy density contrast $\delta(\xvec,t)$ were statistically homogeneous and isotropic, and if it were to evolve according to linear perturbation theory, then $P_m(k)$ would be the power spectrum that's conjugate to the two-point correlation function $\langle \delta(\xvec,t_0) \, \delta(\yvec,t_0) \rangle$ in the Universe today ($t = t_0$). 
For low-$k$ modes that are still in the linear regime today, $k^3 P_m(k)$ measures dark matter inhomogeneities on the length scale $2 \pi / k$.  
For high-$k$ modes that are in the nonlinear regime, $P_m(k)$ provides a convenient benchmark against which different models and measurements can be compared.  
In \LCDM{} Cosmology the linear matter power spectrum is predicted to take the form $P_m = P_\zeta \; T_{m,\zeta}^2$ where $P_\zeta(k)$ is the power spectrum of primordial (adiabatic) curvature perturbations and $T_{m,\zeta}(k)$ is the matter transfer function accounting for the adiabatic mode evolution.  
In a model of \WWDM{} the linear matter power spectrum is modified by two effects:  the ``free-streaming effect'' introduces a scale-dependent modification to the adiabatic component and the ``white-noise effect'' introduces an additional isocurvature component; we discuss these effects further in turn below. 
Taken together the linear matter power spectrum for \WWDM{} may be written as \cite{Amin:2022nlh}\footnote{
The derivations of \erefs{eq:Delta_delta_sq}{eq:Tfs} in \rref{Amin:2022nlh} employ a few simplifying approximations, and thus we interpret the power spectrum as carrying an $O(1)$ theoretical uncertainty.  When we use the power spectrum to derive predictions and constraints, we only quote one significant figure so as to not give the impression of an unduly accurate result. 
} 
\bes{\label{eq:Delta_delta_sq}
    P_m(k) & \approx P_\zeta(k) \; T_{m,\zeta}(k)^2 \; T_\mathrm{fs}(k)^2 + P_{\Scal_\mathrm{dm}}(k) \, T_{m,\Scal_\mathrm{dm}}(k)^2 \\ 
    \Delta_\delta^2(k) & \approx \frac{k^3}{2\pi^2} \; P_\zeta(k) \; T_{m,\zeta}(k)^2 \; T_\mathrm{fs}(k)^2 + \frac{k^3}{2\pi^2} \; P_{\Scal_\mathrm{dm}}(k) \; T_{m,\Scal_\mathrm{dm}}(k)^2 
    \com
}
where $T_\mathrm{fs}(k)$ is the transfer function accounting for the free streaming of \WWDM{}, $P_{\Scal_\mathrm{dm}}(k) = 2\pi^2 / k_\mathrm{wn}^3$ is the (white noise) power spectrum of the additional isocurvature component of \WWDM{}, and $T_{m,\Scal_\mathrm{dm}}(k)^2 = a_0^2 / \aeq^2$ is the corresponding matter transfer function accounting for the isocurvature mode evolution during the matter era.\footnote{\label{footnote5}
Since the isocurvature component of the matter power spectrum is a white noise arising from temporal and spatial fluctuations of the \WWDM{} field, one may not expect these inhomogeneities to grow in the matter era.  However, to be more precise, it is a question of time scales.  If the time scale for the fluctuations is longer than the time scale for the growth of structure (i.e., the fluctuations are slow) then structure would grow. To illustrate this point, consider a region of space of comoving size $\lambda = 2 \pi / k$.  On this length scale, the coherence time of the field fluctuations is $t_\mathrm{coh}(k) \sim a\lambda/v_\ast$ where $v_\ast$ is the typical speed of the field modes with $|{\bm q}| = q_\ast$.  The time scale for growth of structure is $t_\mathrm{grow} \sim \sqrt{G \bar{\rho}} \sim H^{-1}$.  Imposing $t_\mathrm{grow} < t_\mathrm{coh}(k)$, such that  the field fluctuations are ``slow,'' leads to $k < aH/v_\ast$. Note that $aH/v_\ast \sim k_J$ is the usual comoving Jeans wavenumber for gravitational collapse, and it is generally larger than the free-streaming wavenumber $k_\mathrm{fs} = 1 / l_\mathrm{fs}$ by an order-one logarithmic factor \eqref{eq:lfs_t0}. For typical parameters, the Jeans wavenumber today is $a_0 H_0 / v_\ast \sim (3000 \hMpc)(\qast/m/10^{-7})^{-1}$.  To summarize the modes with $k < k_J$ should grow despite the white noise field fluctuations, since the fluctuations on these scales are slow.  
}
This relation assumes that the initial adiabatic and isocurvature components are uncorrelated.
The free-streaming suppression is not expected to impact the isocurvature component~\cite{Ling:2024qfv}, and so the $T_\mathrm{fs}^2$ factor only appears on the adiabatic component.
The isocurvature component of the matter power spectrum survives free streaming, because the isocurvature component is a white noise arising from fluctuations of the \WWDM{} field. 
Free streaming alters these field fluctuations, which correspond to spatially uncorrelated density inhomogeneities (\ie{}, shot noise), but does not alter their statistical properties~\cite{Amin:2025dtd}. 
Instead free streaming removes the correlated density inhomogeneities associated with the usual adiabatic perturbations.  
This behavior is perhaps easier to imagine for particle dark matter, where the white noise component in the power spectrum is associated with the random and independent locations of the individual dark matter particles. 
Although the particles will move around (\ie{}, free stream), the statistics of their spatial distribution will be unchanged, and the white noise component will persist. 
However, for particle dark matter the white noise goes inversely with the particle density ($k_\mathrm{wn} \propto \bar{n}^{1/3}$ rather than $\qast$ for wave dark matter), and the white noise is negligible on observable scales.

Next we discuss the two novel effects of \WWDM{}.  

%=========
\para{Free-streaming effect.}  
Due to the significant velocity dispersion of \WWDM{}, the dark matter free streams an appreciable distance and suppresses the growth of density inhomogeneities on small length scales.  
This free-streaming effect is captured by the transfer function $T_\mathrm{fs}(k)$, which tends to $1$ at small $k$ and tends to $0$ at large $k$.  
The transfer function is well approximated by $T_\mathrm{fs}(k) = \langle \mathrm{sinc}[k \, l_\mathrm{fs}(p,t_0) \bigr] \rangle$ where $l_\mathrm{fs}(p,t)$ is the free streaming length from \eref{eq:lfs_def} and the angled brackets denote averaging with respect to $f(p,t)$ as in \eref{eq:average_def}.  
Using these relations, the free-streaming transfer function can be written as~\cite{Amin:2022nlh}\footnote{
The authors of \rref{Amin:2022nlh} derive $T_\mathrm{fs}(k)$ in terms of $P_\phi(q,t_0)$ as follows.  Using a WKB ansatz they isolate a component of the scalar \WWDM{} field that varies slowly in space and is related to the energy density inhomogeneities.  The field's evolution before radiation-matter equality is governed by the Klein-Gordon equation on a perturbed FRW background with gravitational potentials sourced by the radiation.  They solve for the evolution of the field and its associated density inhomogeneities, which yields the matter power spectrum.  
}
\bes{\label{eq:Tfs}
    T_\mathrm{fs}(k) 
%    & \equiv \bigl< \mathrm{sinc}\bigl[ k \, l_\mathrm{fs}(p,t_0) \bigr] \bigr> \\ 
    & = \frac{\int_0^\infty \! \frac{\dd p}{p} \, \frac{p^3}{2\pi^2} \, f(p,t_0) \ \mathrm{sinc}\bigl[ k \, l_\mathrm{fs}(p,t_0) \bigr]}{\int_0^\infty \! \frac{\dd p}{p} \, \frac{p^3}{2\pi^2} \, f(p,t_0)} 
%    \\ & 
%    = \frac{\int_0^\infty \! \frac{\dd q}{q} \, \frac{q^3}{2\pi^2} \, a(t_0)^3 \, \omega(q,t_0) \, P_{\phi}(q,t_0) \ \mathrm{sinc}\bigl[ k \, l_\mathrm{fs}(q,t_0) \bigr]}{\int_0^\infty \! \frac{\dd q}{q} \, \frac{q^3}{2\pi^2} \, a(t_0)^3 \, \omega(q,t_0) \, P_{\phi}(q,t_0)} \\
%    & = m^2 \, \frac{\int_0^\infty \! \frac{\dd q}{q} \, \frac{q^3}{2\pi^2} \, \frac{\omega(q,t_0)}{m} \, P_{\phi}(q,t_0) \ \mathrm{sinc}\bigl[ k \, l_\mathrm{fs}(q,t_0) \bigr]}{\int_0^\infty \! \frac{\dd q}{q} \, \frac{q^3}{2\pi^2} \, m \omega(q,t_0) \, P_{\phi}(q,t_0)} 
%    \\ & 
    \approx \frac{m^2}{\bar{\rho}(t_0)} \, \int_0^\infty \! \frac{\dd q}{q} \, \frac{q^3}{2\pi^2} \, P_{\phi}(q,t_0) \ \mathrm{sinc}\bigl[ k \, l_\mathrm{fs}(q,t_0) \bigr] 
    \com
}
where the approximation in the equality follows because the modes that dominate the integral are very nonrelativistic $q \approx \qast \ll a(t_0) m$, and consequently $\omega(q,t_0) \approx m$.  
For large-scale modes with small $k$, the sinc function may be approximated as $\mathrm{sinc}(x) = 1 - x^2 / 6 + O(x^4)$, and the integral gives $T_\mathrm{fs}(k) \to 1$ as $k \to 0$.  
For larger values of $k$, the integral initially decreases quadratically, going as $T_\mathrm{fs}(k) \approx 1 - k^2 / 3 \kfs^2$.  
If the power spectrum $P_\phi(q,t_0)$ is sufficiently tightly peaked at $q = \qast$ then the suppression in the power spectrum occurs at 
\bes{\label{eq:kfs_ast}
    \kfsast 
    = |\lvec_\mathrm{fs}(p = \qast)|^{-1} 
    = \biggl( \int_0^{a} \! \frac{\dd a^\prime}{a^\prime} \, \frac{1}{a^\prime \, H(t(a^\prime))} \ \frac{\qast}{\sqrt{\qast^2 + (a^\prime)^2 m^2}} \biggr)^{-1} 
    \per
}
We use this relation to assess the impact of modified cosmological expansion histories on the free streaming of \WWDM{}. 

%=========
\para{White-noise effect.}  
Due to the finite correlation length of the field amplitude in models of \WWDM{}, the energy density contrast develops an isocurvature component with a white-noise power spectrum. 
This can be seen as follows.  
We model the \WWDM{} as a real scalar field $\phi(\xvec,t)$ with a stochastic initial condition that is statistically homogeneous, that is statistically isotropic, that is Gaussian in its Fourier mode amplitudes, and that has two-point correlations given by \eref{eq:P_phi_phi} with power spectrum $P_\phi(|\qvec|,t_i)$.  
The field's stochastic energy density $\rho_\phi(\xvec,t)$ is given by \eref{eq:rho_phi_def}.  
We define the energy density contrast $\delta_\phi$ and the associated power spectrum $P_{\delta}$ as 
\bes{
    \delta_\phi(\xvec,t) = \frac{\rho_\phi(\xvec,t)}{\bar{\rho}(t)} - 1 
    \qquad \text{and} \qquad 
    \langle \delta_\phi(\xvec,t) \, \delta_\phi(\yvec,t) \rangle 
    = \int \! \! \frac{\dd^3 \kvec}{(2\pi)^3} \, P_{\delta}(|\kvec|,t) \, \ee^{\ii \kvec \cdot (\xvec - \yvec)} 
    %= \int_0^\infty \! \frac{\dd k}{k} \, \frac{k^3}{2\pi^2} P_{\delta}(k,t) \, \mathrm{sinc}\bigl( k \, |\xvec - \yvec| \bigr)
    \per
}
Since we have assumed that the stochastic field's fluctuations are Gaussian, the four-point functions reduce to a product of two-point functions.  
In this way, one can show that \cite{Amin:2022nlh}\footnote{This convolution relating $P_\phi$ and $P_\delta$ has appeared in various forms in earlier work studying energy inhomogeneities due to scalar field fluctuations during inflation~\cite{Lyth:1991ub,Liddle:1999pr,Chung:2004nh,Chung:2011xd,Ling:2021zlj}.}
\bes{\label{eq:Pdelta_from_Pphi}
    P_{\delta}(k,t) 
    & = \frac{1}{2 \bar{\rho}(t)^2} \int \! \! \frac{\dd^3 \qvec}{(2\pi)^3} \biggl\{ 
    \Bigl( |\qvec|^2 / a(t)^2 + m^2 \Bigr) \Bigl( |\qvec - \kvec|^2 / a(t)^2 + m^2 \Bigr) 
    \\ & \hspace{3cm}  
    + \Bigl( \qvec \cdot (\qvec - \kvec) / a(t)^2 + m^2 \Bigr)^2 
    \biggr\} \, P_{\phi}(|\qvec|,t) \, P_{\phi}(|\qvec-\kvec|,t) 
    \com
}
which is independent of the orientation of $\kvec$.  
In some models of ultra-light dark matter, the field has a significant amplitude in its $|\qvec| \to 0$ modes.  
However, for \WWDM{} the field is assumed to have a power spectrum $q^3 P_\phi(q,t_i)$ that tends to zero as $q \to 0$ and is peaked at $q = \qast > 0$, as in \eref{eq:rho_phi_def}.  
For asymptotically small $k$, the integral remains dominated by $|\qvec| \approx \qast$, and its value goes to a constant:
\bes{
    \lim_{k \to 0} P_{\delta}(k,t) 
    = \frac{1}{\bar{\rho}(t)^2} \int \! \! \frac{\dd^3 \qvec}{(2\pi)^3} 
    \Bigl( |\qvec|^2 / a(t)^2 + m^2 \Bigr)^2 
    \Bigl( P_{\phi}(|\qvec|,t) \Bigr)^2 
    \per
}
Consequently, the dimensionless power spectrum decreases on large scales: 
\bes{
    \Delta_\delta(k,t) 
    \equiv \frac{k^3}{2\pi^2} P_{\delta}(k,t) 
    \propto k^3 \quad \text{for $k \ll \qast$} 
    \per
}
In other words, the field's finite spatial correlation length $\sim 2 \pi / \qast$ is indistinguishable from point-like spatial correlation (white noise) from the perspective of larger length scales, and the associated white noise power spectrum falls as $k^3$ toward small $k$.  
Using the broken power-law model for the field's power spectrum \eqref{eq:broken_power_law} one may evaluate~\cite{Ling:2024qfv} 
\ba{
    \lim_{k \to 0} \Delta_\delta(k,t) 
    = \underbrace{\frac{\alpha^2 \nu^2}{2(3/2 + \alpha) (\alpha + \nu) (\nu - 3/2)} \biggl( \frac{k}{\qast} \biggr)^3}_{\ \equiv \ (k / k_\mathrm{wn})^3} + O(k^3 / \qast a^2 m^2) 
    \com
}
where we've dropped terms that are negligible when the $\qast$ modes are nonrelativistic.  
For a given choice of $\alpha$ and $\nu$, the value of $\qast$ must not be taken too small, since otherwise the white noise tail of the power spectrum would lead to an enhancement of power, which is not observed at $k \approx (1-10) \hMpc$.  
For reasonable values of $\nu$ and $\alpha$, this can be avoided by taking $\qast \gtrsim 1000 \, \hMpc$.  

%==================================
% Standard expansion
%==================================
\section{Standard expansion}
\label{sec:model_0}

%=========
%Motivation
Before we study free streaming in modified cosmological expansion histories, we begin here by considering the constraints on warm wave dark matter in the Standard Expansion (\SE{}) Cosmology.  
We anticipate that our study will align with the results of earlier work~\cite{Amin:2022nlh,Liu:2024pjg,Ling:2024qfv}.  
In particular \rref{Amin:2022nlh} concluded that warm wave dark matter with $\qast = 1000 \Mpc^{-1} \approx 1500 \hMpc$ needs $m \gtrsim 4 \times 10^{-19} \eV$ in order to avoid an unobserved suppression of power at $k_\mathrm{obs} = 10 \Mpc^{-1} \approx 15 \hMpc$, which is probed by measurements of \Lya{} forest spectra. 
This corresponds to an effective velocity of $v_\ast = q_\ast / m \approx 2 \times 10^{-8}$.  
In this section we revisit these constraints with the goal of extending them into the broader two-dimensional parameter space consisting of \WWDM{} momentum $\qast$ and mass $m$.  

%=================
% --- FIGURE 4 --- 
%=================
\begin{figure}[t!]
\centering
\includegraphics[width=0.7\textwidth]{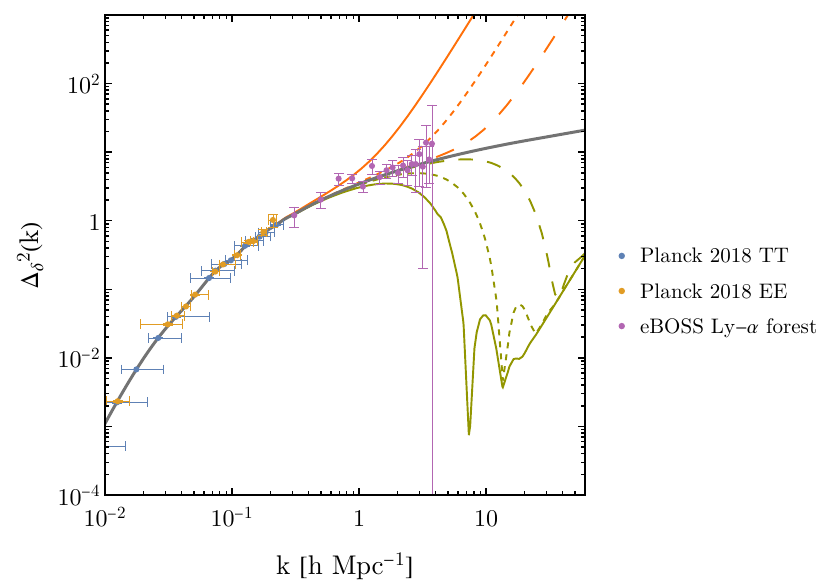}
\caption{\label{fig:model_0_Pm}
Dimensionless linear matter power spectrum.  The various curves show the predictions of \WWDM{}.  Orange lines illustrate model parameters for which the white-noise effect is dominant: $m = 10^{-16} \eV$ and $\qast = 1.8 \times 10^{2} \hMpc$ $= 7.4\times 10^{-28} \,\text{eV}$ (solid), $\qast = 4 \times 10^{2} \hMpc$ $= 1.6\times 10^{-27} \,\text{eV}$ (short-dashed), and $\qast = 10^{3} \hMpc$ $= 4.1\times 10^{-27} \,\text{eV}$ (long-dashed).  The green lines correspond to the model parameters for which the free streaming effect is dominant: $\qast / m = 1.6 \times 10^{-7}$ (solid), $\qast / m = 8.2 \times 10^{-8}$ (short-dashed), and  $\qast / m = 2.7 \times 10^{-8}$ (long-dashed). The data points (taken from \rrefs{Chabanier:2019eai,MacInnis:2024znd}) correspond to measurements of \CMBR{} temperature anisotropies (Planck 2018 TT), \CMBR{} polarization anisotropies (Planck 2018 EE), and \Lya{} forest spectra (eBOSS DR14 Ly-$\alpha$ forest).  The dark gray curve indicates the \LCDM{} prediction that best fits the Planck 2018 data.  
}
\end{figure}

%=========
%Measurements
Various different observations probe dark matter inhomogeneities across a vast range of length scales and at several times in the cosmic history~\cite{Bechtol:2022koa,Drlica-Wagner:2022lbd}.  
On linear and quasi-linear scales ($k \lesssim 1 \hMpc$), one can map these observations into measurements of the late-time ($z = 0$) dimensionless linear matter power spectrum $\Delta_\delta^2(k)$ by assuming a model of cosmology \cite{Tegmark:2002cy}.  
In \fref{fig:model_0_Pm} we illustrate several of these measurements, which were derived by the authors of \rref{Chabanier:2019eai} assuming a best-fit \textit{Planck} 2018 \LCDM{} cosmology.  
Precision measurements of the \CMBR{} temperature and polarization anisotropies \cite{Planck:2018vyg} determine $\Delta_\delta^2(k)$ with percent-level precision at relatively large length scales, corresponding to $k \lesssim 0.1 \hMpc$.  
On smaller length scales, observations of \Lya{} forest spectra \cite{eBOSS:2018qyj} with eBOSS DR14 \cite{eBOSS:2017pfi} provide strong constraints on $\Delta_\delta^2(k)$ up to $k \approx k_\text{Ly-$\alpha$} = 3.74 \hMpc$.  
We will see that these measurements drive our limits on \WWDM{}, since the modifications to $\Delta_\delta^2(k)$ are most significant at small length scales.  
Other observations probe dark matter inhomogeneities at even larger wavenumbers up to $k \approx (10-50) \hMpc$ where the evolution is nonlinear.  
These include additional measurements of \Lya{} forest spectra \cite{Croft:2000hs,Murgia:2017lwo,Murgia:2018now,Cain:2022ehj,Irsic:2023equ}, measurements of the Milky Way satellite population \cite{Nadler:2019zrb,DES:2020fxi}, and measurements of the UV luminosity function of high-redshift galaxies \cite{Sabti:2021unj}.  
We do not account for these measurements in our analysis, since they probe small scales where the mode evolution is highly nonlinear and additional techniques such as $N$-body simulation are required to bridge the gap between theory and data.  
Although it would be interesting to derive predictions for these small-scale observables in a \WWDM{} cosmology, we feel that this calculation is beyond the scope of our work.  
The constraints that we derive from the eBOSS \Lya{} measurements are conservative in the sense that stronger constraints could be derived by using the smaller-scale observations.  
We also consider the projected sensitivity of the proposed CMB-HD experiment~\cite{MacInnis:2024znd}, which aims to achieve sub-percent level precision up to $k \approx 10 \hMpc$ using precision observations of \CMBR{} gravitational lensing.  

%=========
%Methods
To assess the compatibility of \WWDM{} with measurements of the linear matter power spectrum $\Delta_\delta^2(k)$, we perform the following analysis.  
We calculate $\Delta_\delta^2(k)$ using \eref{eq:Delta_delta_sq} for a range of values of the dark matter momentum $\qast$ and the dark matter mass $m$, while holding fixed the \LCDM{} cosmological parameters.  
For each point in the parameter space, we employ the method of least squares to calculate $\chi^2$, which quantifies the deviation between the model and the measurement.  
This approach assumes that all of the data points are uncorrelated, which is not expected to be the case; however, since our limit is mainly driven by only the few data points with largest $k$, we do not expect correlations to significantly degrade the limit.  
We find that $\chi^2$ is minimized at large $\qast$ and small $\qast / m$ where the modifications to $\Delta_\delta^2(k)$ from \WWDM{} at scales inaccessible to the measurements.  
We draw the $90\%$ C.L. curve along the boundary where $\Delta \chi^2 = 4.61$.  
To place constraints, we calculate $\chi^2$ using the eBOSS Ly-$\alpha$ measurements from \fref{fig:model_0_Pm}, and to assess the projected sensitivity of CMB-HD (see below) we use forecasted errors from figure 2 of \rref{MacInnis:2024znd}.

\begin{figure}[t]
\centering
\includegraphics[width=1\textwidth]{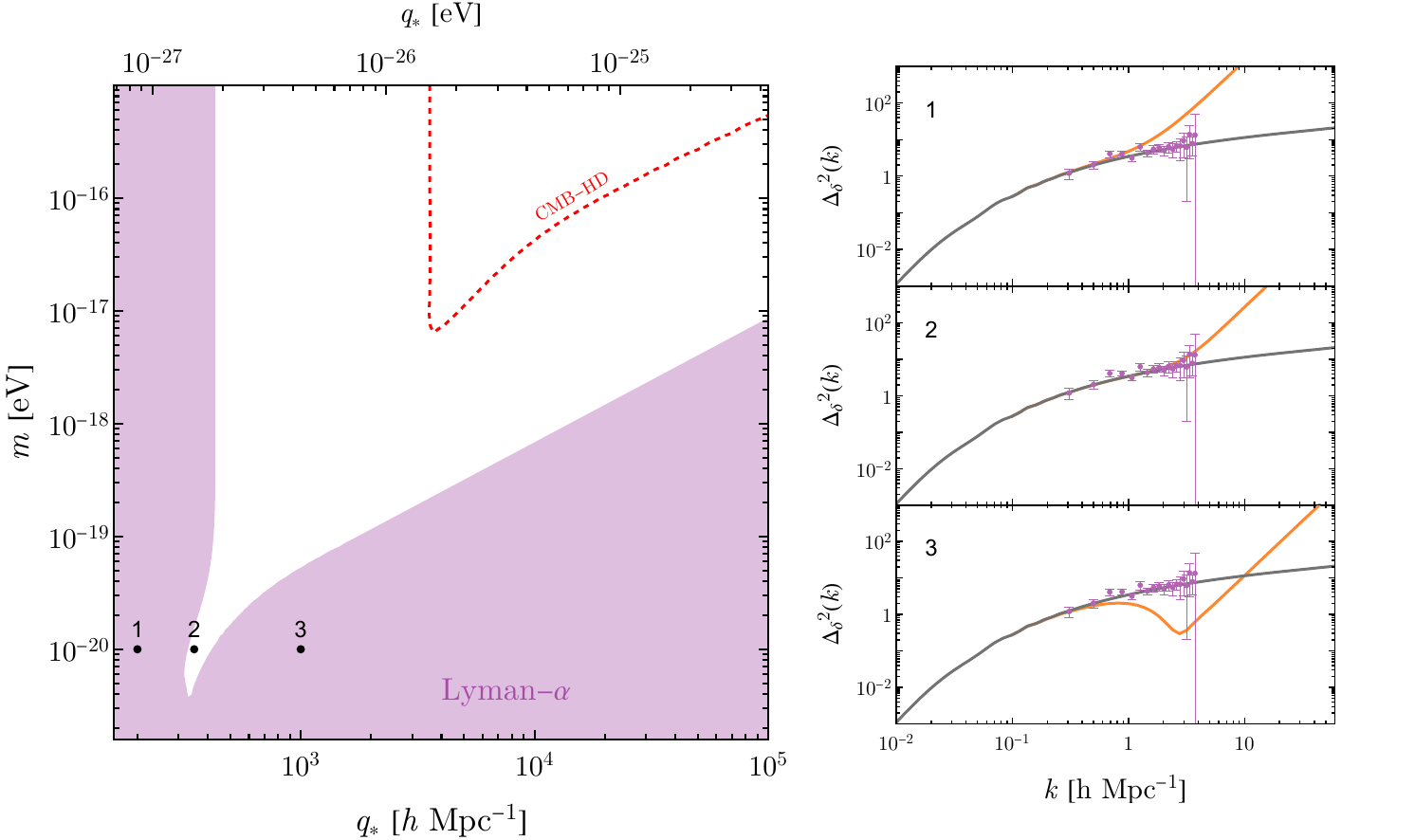}
\caption{\label{fig:model_0_params}
\textit{Left:} Constraints on the \WWDM{} parameter space assuming Standard Expansion Cosmology.  We plot the characteristic dark matter momentum $\qast$ and dark matter mass $m$, and we fix the \LCDM{} parameters; see below \eref{eq:rho_SM_DM_Lambda}.  The purple-shaded region is excluded at the $90\%$ C.L. by eBOSS \Lya{} measurements, which are in good agreement with the predictions of \LCDM{} Cosmology, and the red-dashed curve shows the projected sensitivity of the CMB-HD experiment.  \textit{Right}: Dimensionless power spectrum at three benchmark points in the \WWDM{} parameter space.  We show the \LCDM{} prediction (black), and the \WWDM{} prediction (orange) for $m = 10^{-20} \eV$ and (1) $\qast = 2 \times 10^2 \hMpc$, (2) $\qast = 3.5 \times 10^2 \hMpc$, and (3) $\qast = 10^3 \hMpc$. }
\end{figure}

%=========
%Results
The results of our analysis are presented in \fref{fig:model_0_params}. 
The purple-shaded region of the \WWDM{} parameter space is excluded at the $90\%$ C.L. by eBOSS \Lya{} measurements, which are in good agreement with the predictions of \LCDM{} Cosmology, and the white-shaded region is allowed. 
Toward small $\qast$ the constraint, $\qast \gtrsim 400 \hMpc$, arises from the white-noise effect of \WWDM{}, which enhances the linear matter power spectrum with respect to \LCDM{}.  
Toward large $\qast/m$ the constraint, $\qast / m \lesssim 1.1 \times 10^{22} \hMpc / \mathrm{eV} \approx 5 \times 10^{-8}$, arises from the free-streaming effect of \WWDM{}, which suppresses the linear matter power spectrum with respect to \LCDM{}.  
The Jean's wavenumber (see footnote~\ref{footnote5}) is approximately $k_J \approx 6000 \hMpc$ along the constraint, which is much larger than $k_\Lya{} \approx 4 \hMpc$, indicating that our treatment of isocurvature's evolution is self-consistent. 
In the lower-right corner, the Jeans wavenumber $k_J \approx 1 \hMpc$ becomes as small as $k_\Lya{}$, but at these parameters the power spectrum is strongly suppressed by free-streaming. 
In the lower-left corner (hook-shaped feature), we find that the constraint is slightly weaker than what one would have inferred by considering the white noise and free streaming effects separately.  
Here, the white-noise enhancement is counterbalanced by free-streaming suppression, and the global lower bound on the \WWDM{} dark matter mass is found to be $m > 4 \times 10^{-21} \eV$ at $90\%$ confidence if $\qast \approx 300 \hMpc$.  

%=================
% --- FIGURE 5 --- 
%=================

%=========
The red-dashed curve on \fref{fig:model_0_params} indicates the projected constraint on the \WWDM{} parameter space for the anticipated sensitivity of the CMB-HD experiment~\cite{MacInnis:2024znd}.  
CMB-HD expects to achieve sub-percent precision up to $k \approx 10 \hMpc$ and sub-$10\%$ precision up to $k \approx k_\text{CMB-HD} = 47.3 \hMpc$, which is about a factor of $10$ higher in $k$ than the eBOSS \Lya{} measurements.  
Thus one should expect that the CMB-HD measurements would strengthen the lower limit on $\qast$ and the upper limit on $\qast/m$ each by a factor of about $10$, and that together they would tighten the lower limit on $m$ by a factor of about $100$. 
Our calculation reveals that the CMB-HD projected sensitivity is stronger (by a factor of a few) than these parametric scaling estimates would suggest. 
This is driven by the relatively smaller uncertainties in the CMB-HD projected sensitivity as compared with \Lya{}. 

%=========
%Modeling caveats
We would emphasize that the limits on the \WWDM{} parameters $m$ and $\qast$ shown in \fref{fig:model_0_params} are tied to how we've modeled the linear matter power spectrum with \erefs{eq:Delta_delta_sq}{eq:Tfs}. 
If one were to adopt a different functional form for $P_m(k)$ and repeat this analysis, then the inferred constraints may differ from the ones shown here. 
We have performed a quantitative test by introducing a fudge factor $f$ into the formula for $T_\mathrm{fs}(k)$ through the replacement $l_\mathrm{fs} \to f l_\mathrm{fs}$. 
If $f$ is equal to $2$ or $1/2$, the limit curves in the $(\qast,m)$ plane shift up or down by a factor of approximately $2$. 
One may interpret this spread as a rough gauge of the theoretical uncertainty associated with how the signal is modeled.

%=========
%Conclusion
Measurements of the linear matter power spectrum, particularly eBOSS \Lya{} forest observations at $k \approx 3 \hMpc$, yield strong constrains on the parameter space of \WWDM{}.  
In order to compare our results with earlier work \cite{Amin:2022nlh}, let us focus on $\qast = 1000 \Mpc^{-1} \approx 1500 \hMpc$.  
Here we find a lower limit on the dark matter mass of $m > 0.8 \times 10^{-19} \eV$, which is about a factor of $5$ weaker than the limit found in \cite{Amin:2022nlh}, namely $m > 4 \times 10^{-19} \eV$.  
This mismatch is a consequence our two studies using \Lya{} observations at different wavenumbers; here we use eBOSS \Lya{} measurements that extend up to $k \approx 3 \hMpc$, whereas the authors of \rref{Amin:2022nlh} took $k_\mathrm{obs} = 10 \Mpc^{-1} \approx 15 \hMpc$.  
Since $\kfs$ is linear in $m$, the factor of $5$ difference in the observable \Lya{} scale translates to a factor of $5$ difference in the \WWDM{} mass limit.  
We reiterate that our constraint is conservative in the sense that we do not account for observations that probe dark matter inhomogeneities on nonlinear scales, which could be used to derive a stronger limit.  
Our work extends earlier work into the broader two-dimensional \WWDM{} parameter space, as we have allowed both $\qast$ and $m$ to vary.  
This approach allows us to identify a corner of the parameter space where the limit is weakened to $m > 4 \times 10^{-21} \eV$ due to a competition of the white-noise effect and the free-streaming effect.  
If an experiment with the sensitivity of CMB-HD did not detect any deviation from the \LCDM{} prediction for the linear matter power spectrum, then we estimate that the lower limit on the dark matter mass would strengthen to $m > 6 \times 10^{-18} \eV$.  
Future observations of large scale structure \cite{Chung:2016wvv,Chung:2023syw} and 21 cm intensity mapping \cite{deKruijf:2024voc} are expected to provide powerful probes of blue-tilted dark matter isocurvature.  

%==================================
% Early matter domination
%==================================
\section{Early matter domination}
\label{sec:model_1}

%=========
%Motivation
As a first example of a modified cosmological expansion history, we consider early matter domination (\EMD{})~\cite{Giudice:2000ex,Erickcek:2011us,Waldstein:2016blt,Drees:2017iod,StenDelos:2019xdk,Erickcek:2020wzd}.  
In this framework, one assumes that the cosmological energy budget is extended to include a new energy component that is nonrelativistic, temporarily dominates the energy budget, and subsequently decays to relativistic Standard Model particles.  
Details regarding the properties and interactions of the new particle species are not directly relevant to this analysis, and many particle physics embeddings have been studied. 
For example, the new energy component may correspond to a population of string moduli that decay slowly via Planck-suppressed operators~\cite{Coughlan:1983ci,Banks:1993en,deCarlos:1993wie,Nakamura:2006uc,Bodeker:2006ij}.  
In order to avoid strong constraints from neutrino thermalization and nucleosynthesis~\cite{Holtmann:1998gd,Cyburt:2002uv,Kawasaki:2004qu,Kawasaki:2008qe,Kawasaki:2017bqm,Alves:2023jlo}, the unstable relic must complete its decay before the start of Big Bang Nucleosynthesis (\BBN{}).  
As a fiducial reference point, we use the time when the plasma temperature was $T \approx 5 \MeV$ and the \FRW{} scale factor was $a_\BBN{} / a_0 \approx 3.4 \times 10^{-11}$~\cite{deSalas:2015glj}.  
The essential idea of our study is that during the \EMD{} epoch, the Hubble expansion rate $H(t)$ is smaller than it would be in a \SE{} Cosmology and the free streaming length of warm wave dark matter is correspondingly larger, which then threatens to run afoul of constraints on the linear matter power spectrum if the dark matter mass is too small.  

%=========
%Modeling
The cosmological expansion rate $H(t)$ is given by the first Friedmann equation \pref{eq:Friedmann_eqn}, which we reproduce here: 
\ba{
    \tfrac{\dd}{\dd t} a(t) = a(t) \, H(t) 
    \qquad \text{where} \qquad 
    H(t) = \bigl[ \rho_\SM{}(t) + \rho_\DM{}(t) + \rho_\DE{}(t) + \rho_\X{}(t) \bigr]^{1/2} / \sqrt{3} \Mpl
    \per
}
The energy densities carried by the Standard Model ($\rho_\SM{}$), by the dark matter ($\rho_\DM{}$), and by the dark energy ($\rho_\DE{}$) are given by \eref{eq:rho_SM_DM_Lambda} where $T(t)$ is the Standard Model plasma temperature at time $t$.  
We assume that there is an energy transfer from the $\X{}$ sector into the Standard Model, which can be described by the following system of kinetic equations: 
\bsa{eq:model1_kinetic_eqns}{
    \tfrac{\dd}{\dd t} \rho_\SM{}(t) + 4 H(t) \, \rho_\SM{}(t) & = + \Gamma_\X{} \, \rho_\X{}(t) \\ 
    \tfrac{\dd}{\dd t} \rho_\X{}(t) + 3 H(t) \, \rho_\X{}(t) & = - \Gamma_\X{} \, \rho_\X{}(t) 
    \per
}
Here $\Gamma_\X{}$ parametrizes the rate at which energy is transferred from $\X{}$ to the Standard Model plasma ($\SM{}$); in a particle physics model for this interaction, $\Gamma_\X{}$ would correspond to the decay rate of $\X{}$ particles into relativistic Standard Model particles.  
In order to solve these first-order differential equations, it is necessary to specify corresponding boundary conditions.  
For the \LCDM{} energy components we impose ``initial'' conditions today, denoting the age of the universe today by $t_0$: 
\ba{
    a(t_0) = a_0 
    \ , \quad 
    \rho_\SM{}(t_0) = \Omega_\SM{} \, \rho_\mathrm{crit} 
    \ , \quad 
    \rho_\DM{}(t_0) = \Omega_\DM{} \, \rho_\mathrm{crit} 
    \ , \quad 
    \rho_\DE{}(t_0) = \Omega_\Lambda \, \rho_\mathrm{crit} 
    \com
}
where $\rho_\mathrm{crit} \equiv 3 \Mpl^2 H_0^2$ is the cosmological critical energy density today, where $H_0 = H_{100} h$ is the Hubble constant, and where $H_{100} = 100 \km / \mathrm{sec} / \mathrm{Mpc}$.  
For the $\X{}$ component, we impose an initial condition at an early time, denoted by $t_i$:
\ba{
    \rho_\X{}(t_i) = m_\X{} \, a(t_i)^{-3} \, \Ncal_{X,i}
    \per
}
Here we have introduced $m_\X{}$, which corresponds to the mass of the $\X{}$ particles, as well as $\Ncal_{X,i} = (a^3 n_\X{})_i$, which corresponds to the initial comoving number density of $\X{}$ particles.   
The $\X{}$ energy component introduces an additional $3$ parameters: $m_\X{}$, $\Gamma_\X{}$, $\Ncal_{X,i}$, and the cosmological expansion is insensitive to the value of $t_i$ provided that $0 < t_i \ll \Gamma_\X{}^{-1}$ since the comoving number density is approximately conserved before $\X{}$ decays.  
It is useful to identify $\anr = \qast / m$, which corresponds to the value of the FRW scale factor when the dark matter particles (Fourier modes) with comoving momentum (wavenumber) of $p_\ast$ ($= \qast$) become nonrelativistic. 
It is also convenient to define $a_\mathrm{end}$ to be the value of the FRW scale factor when the age of the universe is comparable to the decay rate of $\X{}$ particles: $t_\mathrm{end} \approx H(t_\mathrm{end})^{-1} \approx \Gamma_\X{}^{-1}$
We solve the coupled system of kinetic equations along with the boundary conditions and the first Friedmann equation in order to calculate the cosmological expansion rate $H(t)$ and its impact on the free streaming  wavenumber of warm wave dark matter $\kfsast$ through \eref{eq:kfs_ast}. 

%=================
% --- FIGURE 6 --- 
%=================
\begin{figure}[t]
\centering
\includegraphics[width=0.48\textwidth]{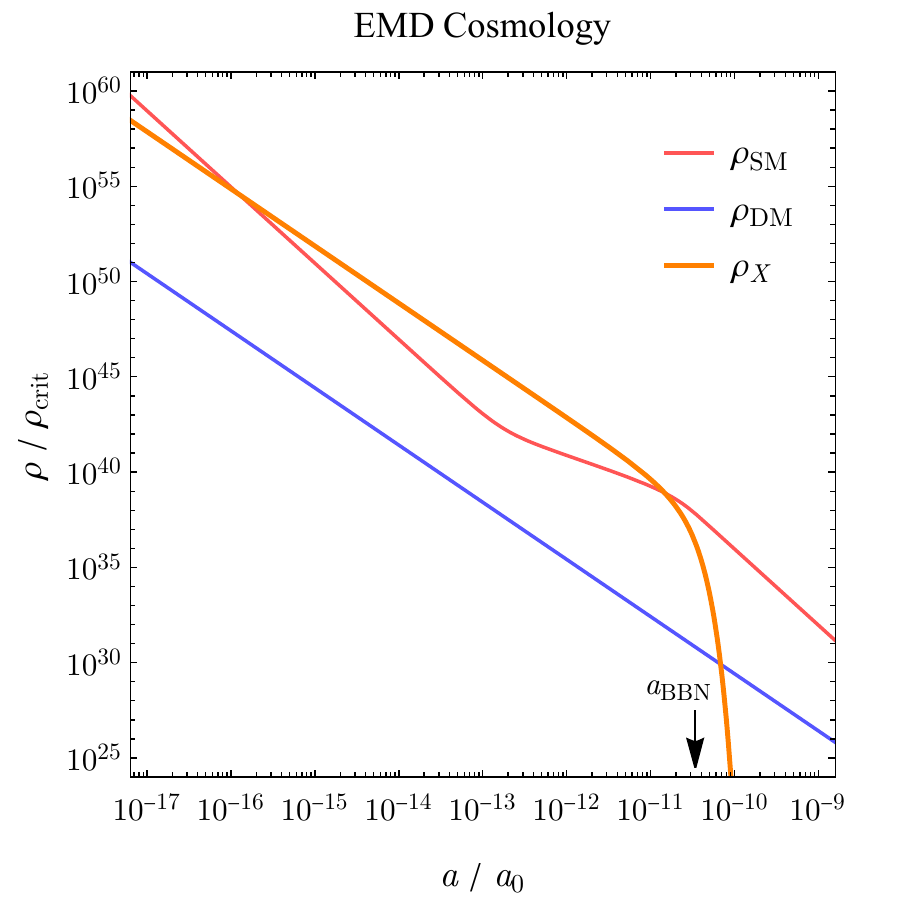} \hfill 
\includegraphics[width=0.48\textwidth]{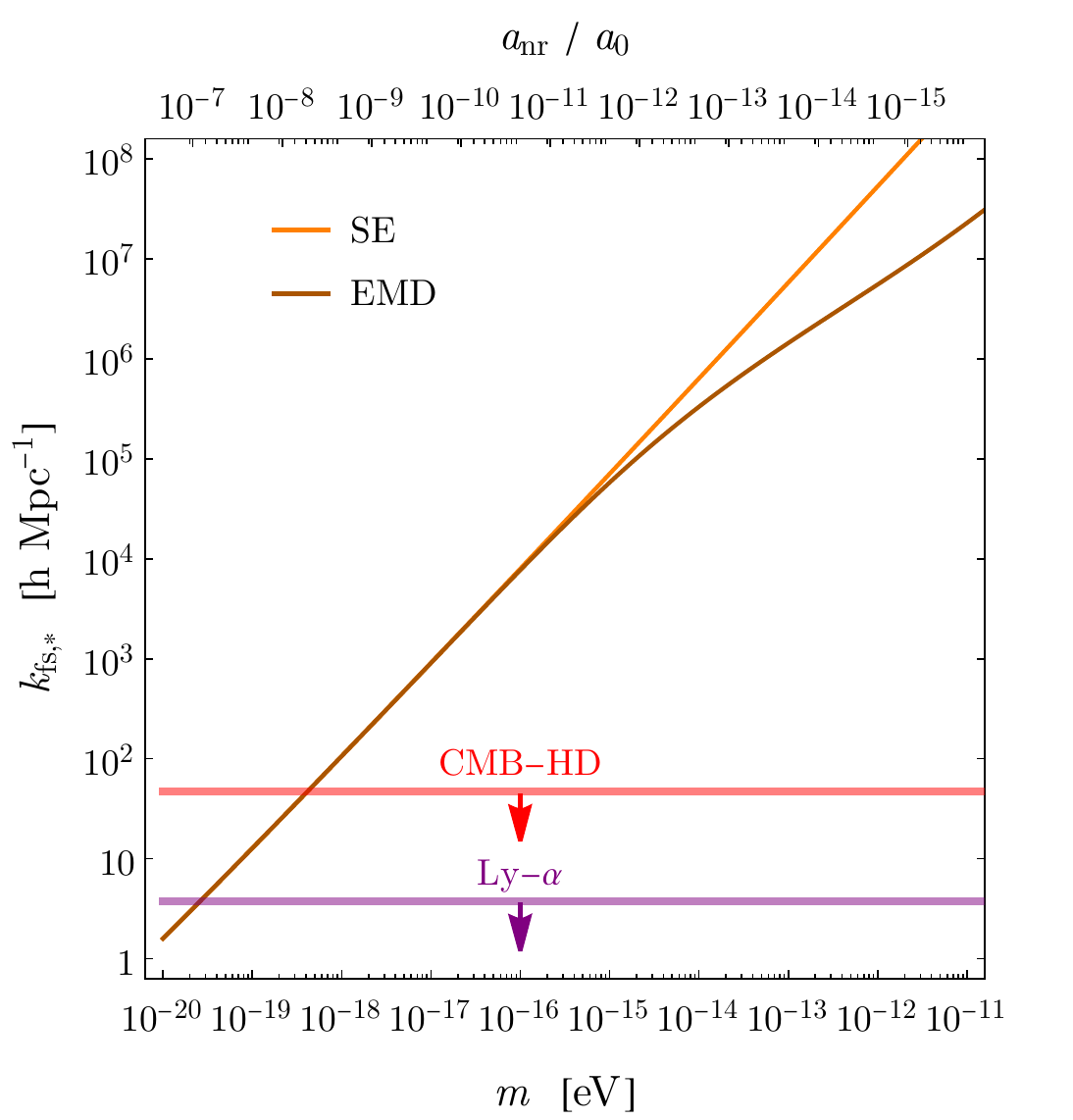} 
\caption{\label{fig:model_1}
\textit{Left:}  Evolution of the various energy components in the \EMD{} Cosmology.  We plot the energy density $\rho$ divided by critical energy density today $\rho_\mathrm{crit} = 3 M_\mathrm{pl}^2 H_0^2$ as a function of scale factor $a$ divided by scale factor today $a_0$.  During the \EMD{} era ($10^{-16} \lesssim a/a_0 \lesssim 10^{-11}$), the energy density carried by the $\X{}$ component ($\rho_\X{}$, orange) dominates over the Standard Model component ($\rho_\SM{}$, red), and the dark matter component ($\rho_\DM{}$, blue), and subsequently it decays away transferring its entropy to the Standard Model.  This entropy injection must complete before the start of \BBN{} and neutrino decoupling, indicated by $a_\BBN{} / a_0 \approx 3.4 \times 10^{-11}$.  To generate this example, we have taken $m_\X{} = 10 \GeV$, $\Gamma_\X{} = 10^{-22} \GeV$, and $\mathcal{N}_{\X{},i} = 2.7 \times 10^{-14}$.  
\textit{Right:}  Predicted free streaming  wavenumber $\kfsast$ in a model of \WWDM{}, comparing the two cosmologies: \SE{} (orange) and \EMD{} (brown).  For the \WWDM{} parameters we take $\qast = 500 \hMpc$ and vary the dark matter mass $m$.  For this \EMD{} parameters, we use the same values as in the left panel.  The horizontal bands indicate to the largest wavenumbers probed by \Lya{} and CMB-HD observations.  For each value of $m$, the top axis shows $\anr = \qast / m$, which is the scale factor when the modes with comoving wavenumber $\qast$ become nonrelativistic. 
}
\end{figure}

%=========
%Results
In the left panel of \fref{fig:model_1} we show the time evolution the various cosmological energy components for a fiducial choice of parameters.  
The $\X{}$ component carries a comparable energy to the radiation at $a / a_0 \approx 10^{-16}$, and it remains the dominant component until $a / a_0 \approx 10^{-11}$ when it decays and transfers its entropy to the radiation.  
Since the radiation is ``heated'' by this entropy injection~\cite{Giudice:2000ex}, the universe must expand longer in bring the radiation temperature down to the observed \CMBR{} temperature.  
As a result, the energy density during the \EMD{} is \textit{smaller} than it would be in an \SE{} Cosmology, the Hubble parameter $H$ is smaller, the free streaming length $\lfs$ is larger, and $\kfsast$ is smaller.\footnote{Note that we hold fixed $q_\ast$ when we compare the standard and modified cosmologies.  The authors of \rref{Gelmini:2008sh} study thermal relic dark matter in a model where kinetic decoupling occurs during a phase of early matter domination.  The \EMD{} phase leads to earlier decoupling, a longer period of redshifting, and colder dark matter with larger $\kfs$.  We would obtain similar results if we fixed the initial physical momentum (rather than comoving) when comparing cosmologies. 
%This result contrasts with \cite{Gelmini_2008}, where the free-streaming scale was studied after kinetic decoupling. 
%In our work, the velocity dispersion of WWDM is determined by the peaked momentum $q_*$ and the dark matter mass $m$, which are free parameters. Consequently, the velocity dispersion is not dictated by the cosmological history. 
}
This modification to the free streaming  wavenumber $\kfs$ is illustrated on the right panel of \fref{fig:model_1}.  
For the parameters chosen to make this figure, the \EMD{} ends at $a_\mathrm{end} / a_0 \approx 10^{-11}$. 
For smaller values of the dark matter mass $m$, the dark matter remains relativistic for a longer time.  
In particular if $m = 2 \times 10^{-15} \eV$ then \WWDM{} with momentum $\qast = 500 \hMpc$ becomes nonrelativistic at $\anr = \qast/m \approx 10^{-11} a_0$, which is comparable to $a_\mathrm{end}$.  
Lighter dark matter becomes nonrelativistic later, meaning that it remains relativistic throughout the \EMD{}, and consequently the free-streaming  wavenumber is unaffected by the modified cosmological expansion history (orange and brown curves overlap).  
For larger values of $m$, the dark matter becomes nonrelativistic earlier, during the \EMD{} and the free-streaming wavenumber $\kfs$ is reduced.  
This reduction in $\kfs$ can be strong for sufficiently large mass $m$, but then the free streaming is moved to tiny length scales that are observationally inaccessible. 

%=========
%Conclusion
We find that an early matter-dominated era prior to the epoch of nucleosynthesis would tend to reduce the free streaming  wavenumber $\kfs$ of \WWDM{}.
This effect can be appreciable (an order one factor or more) if the dark matter mass $m$ is large enough to ensure that dark matter with momentum $\qast$ becomes nonrelativistic before the end of the \EMD{}.  
An \EMD{} epoch would also modify the transfer function describing the evolution of the adiabatic perturbations, which provides an opportunity for complementary probes of \WWDM{}. 
However, since the \EMD{} must end before \BBN{}, the mass needed for a significant modification to free streaming is $m \gtrsim 2 \times 10^{-15} \eV$ for $\qast = 500 \hMpc$, corresponding to a very large $\kfs \approx 10^5 \hMpc$, which would be extremely challenging to observe.  

%==================================
% Early dark energy
%==================================
\section{Early dark energy}
\label{sec:model_2}

%=========
%Motivation
Measurements of the Hubble constant $H_0$ using the supernova distance ladder~\cite{Riess:2020fzl} reveal an approximately $10\%$ mismatch with the prediction for $H_0$ in \LCDM{} Cosmology when the cosmological parameters are fit to the CMB measurements~\cite{Planck:2018vyg}.  
This discrepancy, which is generally known as ``the Hubble tension,'' has sparked the exploration for alternative cosmological models in which the predicted $H_0$ can be reconciled with the direct measurements~\cite{DiValentino:2021izs}.  
One of the most well studied models, known as ``early dark energy'' (\EDE{}), introduces an additional component that behaves like dark energy until the epoch of radiation-matter equality (time $t = t_\mathrm{eq}$), at which time it composes as much as $5\%$ of the cosmological energy budget, and soon afterward it decays away quickly~\cite{Karwal:2016vyq,Poulin:2018cxd,Smith:2019ihp,Murgia:2020ryi}.  
In this section we assess the impact of early dark energy on the free streaming of warm wave dark matter, and we investigate the extent to which measurements of the small-scale dark matter power spectrum can probe \EDE{}.  
Earlier work \cite{Kreisch:2019yzn} has investigated the connection between neutrino free streaming and cosmological tensions, including the $H_0$ tension.  

%=========
%Modeling
We extend the cosmological energy budget to include an additional energy component $\X{}$ that behaves like a cosmological constant at early time and that decays away quickly at around radiation-matter equality.  
We model the energy density in the $\X{}$ component as~\cite{Poulin:2018cxd} 
\bes{\label{eq:model2_rhoX}
    \rho_\X{}(t) = 
    \frac{2}{(a / a_c)^{3 (w_n + 1)} + 1} \, f_\EDE{}(t_c) \, \rho_\mathrm{tot}(t_c) 
    \com
}
where $w_n = (n-1)/(n+1)$ and where $a_c = a(t_c)$.  
This evolution would arise as the average behavior of a homogeneous scalar field oscillating in the potential $V(\varphi) \propto (1 - \cos \varphi/f)^n$.  
We adopt the best-fit model of \rref{Poulin:2018cxd}, which has $n = 3$, $a_c = 10^{-3.737} \approx 1.8 \times 10^{-4}$, and $f_\EDE{}(t_c) = 0.050$.  
For this choice of $\rho_\X{}(t)$, we use the formulas in \sref{sec:WWDM} to calculate the cosmological expansion history, and we use \eref{eq:kfs_ast} to calculate the dark matter free streaming  wavenumber $\kfsast$.

%=================
% --- FIGURE 7 --- 
%=================
\begin{figure}[t]
\centering
\includegraphics[width=0.49\textwidth]{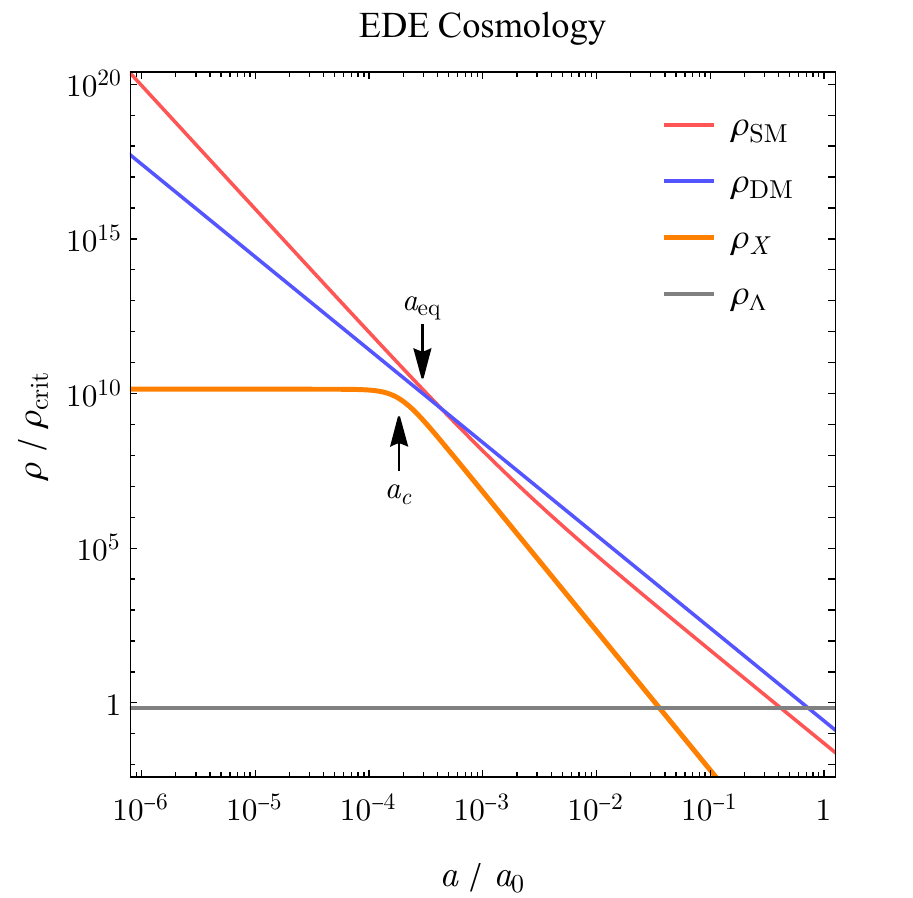} \hfill 
\includegraphics[width=0.49\textwidth]{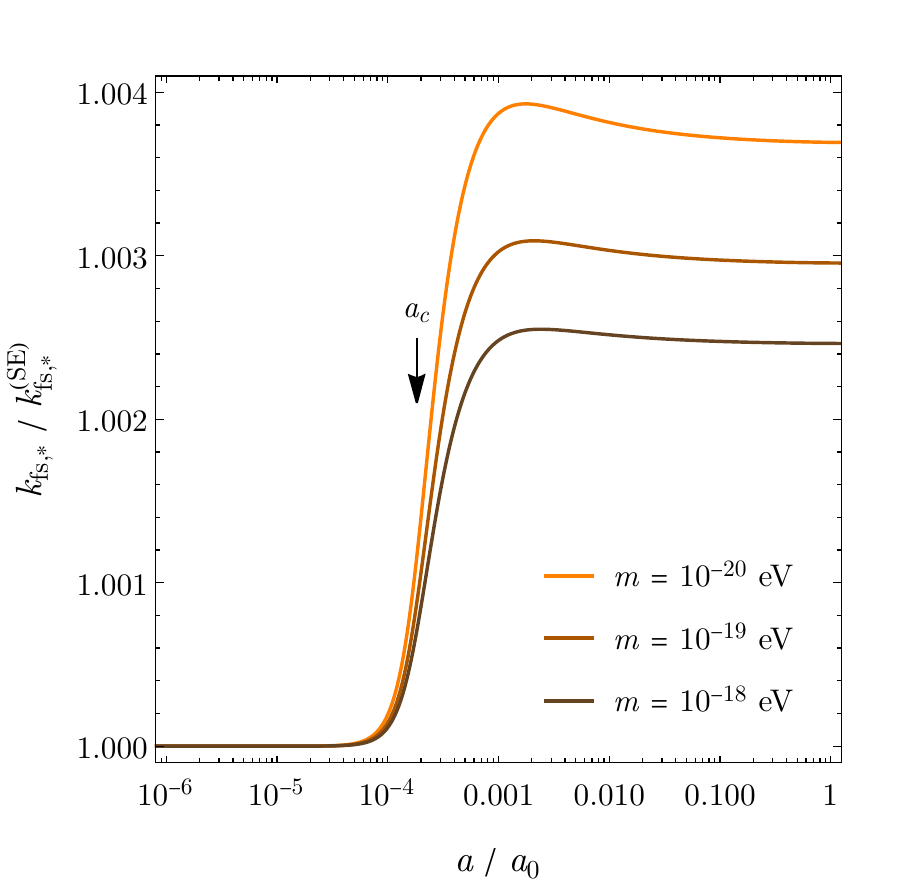}
\caption{\label{fig:model_2}
\textit{Left:} Evolution of the various energy components in the \EDE{} Cosmology.  The \EDE{} component ($\rho_\X{}$, orange) behaves like a cosmological constant until $a_c = 10^{-3.737}$ when it makes up $f_\EDE{} = 5\%$ of the total energy density, and subsequently it redshifts away faster than radiation with equation of state $w_n = 1/2$.  We also indicate radiation-matter equality ($z_\mathrm{eq} = 3387$, $\aeq \approx 3.0 \times 10^{-4}$), which is slightly offset from the crossing of the red and blue lines, since $\rho_\SM{} = \rho_\mathrm{b} + \rho_\mathrm{r}$ contains both baryons and photons. 
\textit{Right:} Modification to the free streaming  wavenumber.  We plot the ratio of $\kfsast$ in the \EDE{} Cosmology with respect to $\kfsast^{(\SE{})}$ in \SE{} Cosmology.  We fix $\qast = 500 \hMpc$ and show three values of $m$ for the sake of illustration, but $\kfsast$ only depends on the ratio $\qast / m$.   
}
\end{figure}

%=========
%Results
The evolution of $\rho_\X{}(t)$ is illustrated in the left panel of \fref{fig:model_2}.  
At early time ($a < a_c$) the energy density carried by $\X{}$ is static, behaving like a cosmological constant.  
At late time ($a > a_c$) the energy decreases to zero with $w_n = 1/2$.   
The right panel of \fref{fig:model_2} shows the time evolution of the free streaming  wavenumber $\kfsast(a)$ for the modes with $q = \qast$.  
For each of the curves we take $\qast = 500 \hMpc$, and we vary the dark matter mass $m$ across the three curves.  
For all three masses shown, the free streaming  wavenumber begins to grow with respect to \SE{} Cosmology (\ie{}, $\kfsast / \kfsast^{(\SE{})} > 1$) at $a_c = 10^{-3.737}$ where the \EDE{} energy density is most significant.  
The \EDE{} increases the energy density by about $5\%$, which increases the Hubble parameter by about $2.5\%$, which reduces the free streaming length scale $\lfs$ (and increases the free streaming wavenumber $\kfsast$) by a smaller factor due to the time integration in \eref{eq:lfs_def}. 
Hence the effect seen in \fref{fig:model_2} is at the level of $0.2\%$ to $0.4\%$.  
This enhancement grows larger for smaller values of the dark matter mass $m$.  
The largest enhancement is achieved at masses for which the dark matter particles become nonrelativistic at around $a_c$.  

%=========
%Conclusion
As expected the free streaming wavenumber $\kfsast$ is less than a percent larger in the \EDE{} cosmology as compared with \SE{} Cosmology.  
In order to reveal this subtle effect, one would require precision probes of the linear matter power spectrum at  wavenumbers $k \approx \mathrm{few} \hMpc$.  
Existing measurements using the \Lya{} forest show no sign of free streaming suppression, and even if they had, then with the current level of uncertainties it would be impossible to distinguish \EDE{} from \SE{} Cosmology.  
Future observations with a sensitivity comparable to CMB-HD, which projects a precision of a few tenths of a percent at these scales (compare with \fref{fig:model_0_Pm}), may be able to discriminate between \EDE{} and \SE{}.

%==================================
% Very early dark energy
%==================================
\section{Very early dark energy}
\label{sec:model_3}

%=========
%Motivation
Whereas an early dark energy can only make up about $5\%$ of the cosmological energy budget if it becomes relevant around the time of recombination (see \sref{sec:model_2}), the universe could contain much more dark energy at earlier times.  
Prior to nucleosynthesis, an additional dark energy component to the cosmological energy budget would be almost entirely unconstrained.  
However, there is also some room for additional dark energy between the epoch of nucleosynthesis and recombination. 
This idea has been explored in a recent paper~\cite{Sobotka:2024ixo} where it has been called ``very early dark energy'' (\vEDE{}).  
In this section we assess the corresponding impact on the free streaming of warm wave dark matter.  

%=========
%Modeling
We adopt a simplified picture to model the \vEDE{} component $\X{}$.  
Whereas \rref{Sobotka:2024ixo} assumed that the \vEDE{} corresponded to a real scalar field $\varphi$ evolving in the potential $V(\varphi) = m^2 f^2 (1 - \cos\varphi/f)^n$ with $n=8$, we will instead approximate this energy component with a smoothed broken power law.  
To that end, we write the energy density in the $\X{}$ component as 
\bes{\label{eq:model3_rhoX}
    \rho_\X{}(t) = 
    \frac{2}{(a / a_c)^{3 (w_\vEDE{} + 1)} + 1} \, f_\vEDE{}(t_c) \, \rho_\mathrm{tot}(t_c) 
    \com
}
where the three model parameters are the energy fraction $f_\vEDE{}(t_c)$, the scale factor $a_c$, and the equation of state $w_\vEDE{}$.  
In general the parameters can take values $0 \leq f_\vEDE{}(t_c) \leq 1$, $w_\vEDE{} > 1/3$, and $a_\BBN{} < a_c < \aeq$ where $a_\BBN{} \approx 3.4 \times 10^{-11} \, a_0$ and $\aeq \approx 3.0 \times 10^{-4} \, a_0$.  
However, we focus on $f_\vEDE{}(t_c) = 0.98$, $w_\vEDE{} = 1$, and $10^{-8} < a_c/a_0 < 10^{-6}$.  
These values are informed by the careful numerical study present in \rref{Sobotka:2024ixo}.\footnote{
A homogeneous scalar field $\varphi(t)$ in a power-law potential $V(\varphi) \propto |\varphi|^{2n}$ has an effective equation of state $w_\mathrm{eff} = (\tfrac{1}{2} \dot{\varphi}^2 - V) / (\tfrac{1}{2} \dot{\varphi}^2 + V)$.  As $\varphi(t)$ oscillates, $w_\mathrm{eff} \approx 1$ when kinetic energy dominates and $w_\mathrm{eff} \approx -1$ when potential energy dominates, such that $w_\mathrm{eff} \approx w_n = (n-1)/(n+1)$ on average. The $(1-\cos \varphi/f)^8$ potential of \rref{Sobotka:2024ixo} corresponds to $w_8 = 7/9$, but since the effect on free streaming is controlled by the first oscillation's kinetic-dominated phase, we take $w = 1$ in our modeling. } 
They concluded that the \vEDE{} energy density could be as large as $R = 60$ times the energy density in the Standard Model plasma ($f_\vEDE{} = R / (1+R) \approx 0.98$) as long as it dominates well after \BBN{} and well before the epoch of baryon acoustic oscillations and recombination.  
For this choice of $\rho_\X{}(t)$, we use the formulas in \sref{sec:WWDM} to calculate the cosmological expansion history, and we use \eref{eq:kfs_ast} to calculate the dark matter free streaming  wavenumber $\kfsast$.

%=================
% --- FIGURE 8 --- 
%=================
\begin{figure}[t]
\includegraphics[width=0.48\textwidth]{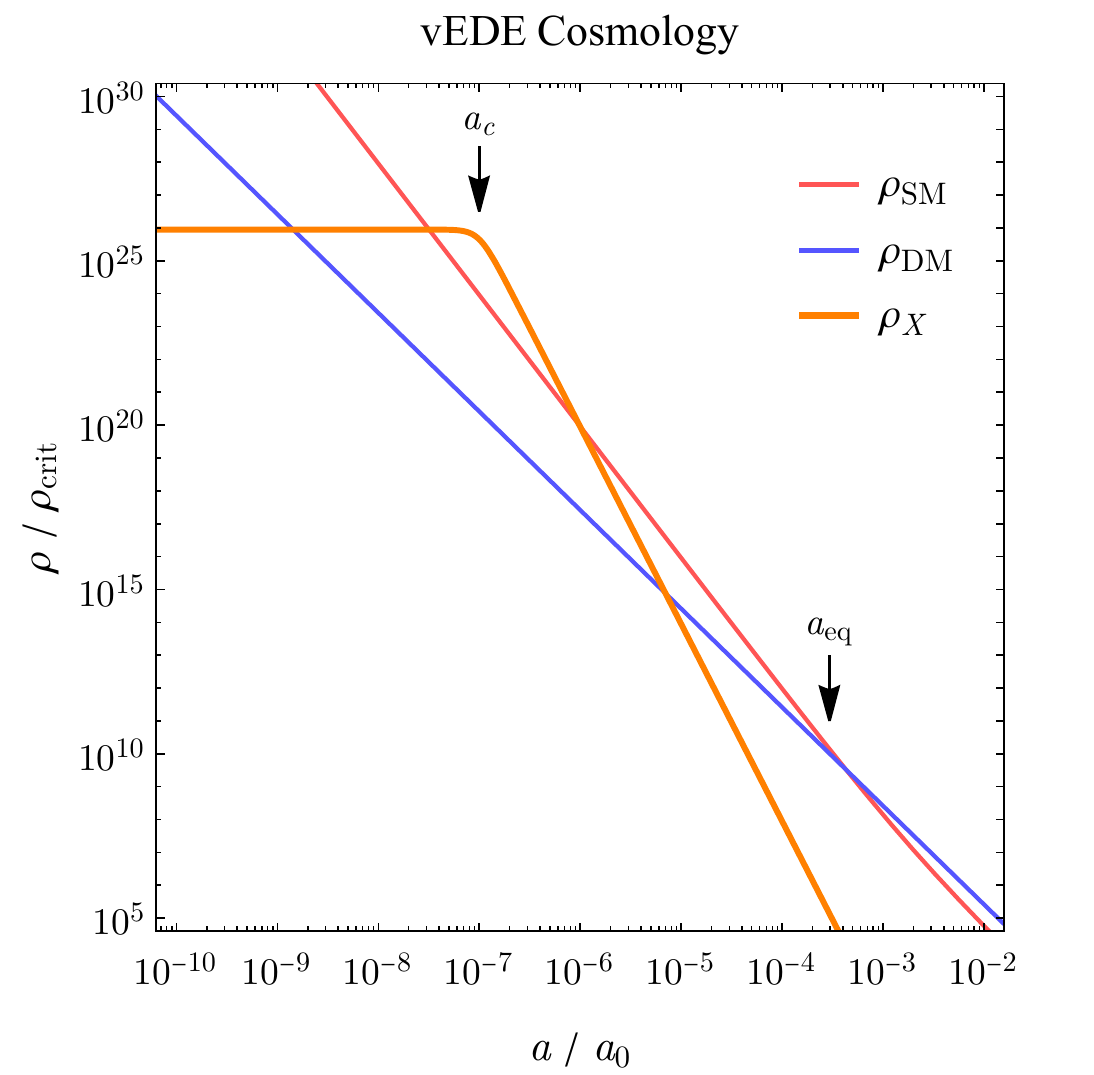} \hfill 
\includegraphics[width=0.55\textwidth]{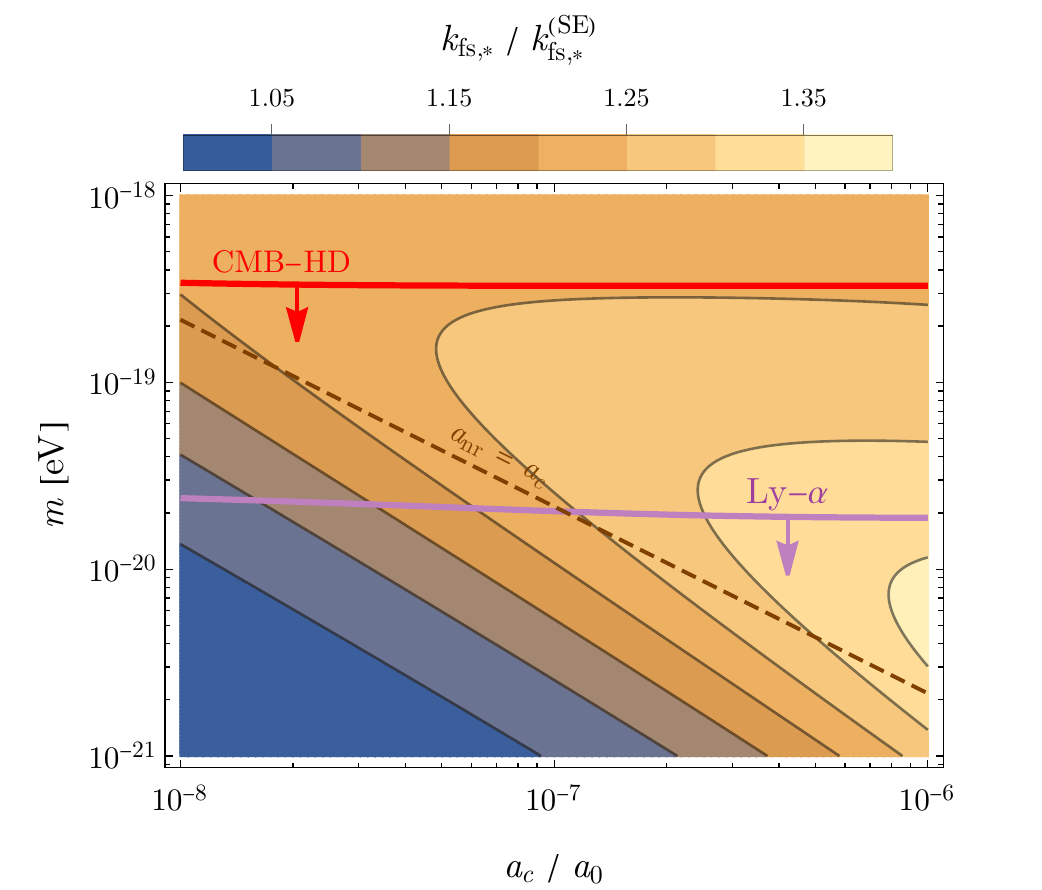} 
\caption{\label{fig:model_3}
\textit{Left:}  Evolution of the various energy components in the \vEDE{} Cosmology.  The \vEDE{} component ($\rho_\X{}$, orange) behaves like a cosmological constant until $a_c = 10^{-7}$ when it makes up approximately $60$ times more energy density than the other species ($f_\vEDE{} = 98\%$), and subsequently it redshifts away faster than radiation with equation of state $w_\vEDE{} = 1$.  
\textit{Right:}  Modification to the free streaming  wavenumber.  Contours show the ratio of $\kfsast$ in the \vEDE{} Cosmology with respect to $\kfsast^{(\SE{})}$ in \SE{} Cosmology.  We fix $\qast = 500 \hMpc$ and show a range of values for $m$, but $\kfsast$ only depends on the ratio $\qast / m$.  The nearly-horizontal lines roughly indicate the reach of \Lya{} and CMB-HD observations to probe this parameter space; \ie. $\kfsast \leq k_\text{Ly-$\alpha$} = 3.74 \hMpc$ and $k_\text{CMB-HD} = 47.3 \hMpc$, respectively.  
Along the brown-dashed line the $\qast$ mode becomes nonrelativistic at $\anr = \qast / m = a_c$.  
}
\end{figure}

%=========
%Results
The evolution of the energy components in the \vEDE{} Cosmology is shown on the left panel of \fref{fig:model_3}.  
Our power-law model \eqref{eq:model3_rhoX} doesn't capture a sub-leading oscillatory behavior that appears in the energy density associated with the cosine potential \cite{Sobotka:2024ixo}, but this behavior is not expected to significantly impact the free-streaming scale.  
The right panel of \fref{fig:model_3} shows the ratio of the free streaming  wavenumber $\kfs$ in the \vEDE{} Cosmology as compared with the \SE{} Cosmology.  
We vary $a_c$, which controls when the dark energy decays, and we vary the dark matter mass $m$.  
The figure shows that $\kfsast$ is larger in the \vEDE{} Cosmology, corresponding to $\kfsast / \kfsast^{(\SE{})} > 1$.  
This enhancement grows with increasing $a_c$ and reaches approximately $40\%$ for $a_c = 10^{-6} a_0$.  
A larger value of $a_c$ would come into conflict with \CMBR{} observables \cite{Sobotka:2024ixo}. 
The parametric scaling changes depending on whether the dark matter becomes nonrelativistic before the dark energy begins to decay ($\anr < a_c$) or afterward ($\anr > a_c$).  
For example, in the lower-left corner, the enhancement is negligible ($\kfsast / \kfsast^{(\SE{})} \approx 1$), because the \WWDM{} becomes nonrelativistic after the \vEDE{} epoch is over ($\anr \gg a_c$).  
In the upper-right corner, the enhancement becomes insensitive to $a_c$, because the \vEDE{} epoch occurs while the \WWDM{} is already nonrelativistic ($\anr \ll a_c$).  
The nearly-horizontal lines roughly indicate the reach of \Lya{} and CMB-HD in this parameter space. 
To draw these curves, we compare the predicted $\kfsast$ with $k_\text{Ly-$\alpha$} = 3.74 \hMpc$ and $k_\text{CMB-HD} = 47.3 \hMpc$ respectively.  

%=========
%Conclusion
Since the \vEDE{} Cosmology allows a much larger energy density in the additional \X{} component, the associated impact on the free streaming of warm wave dark matter is also expected to be larger in comparison with \EDE{} Cosmology.   
This expectation is borne out by the numerical results.  
We find that $\kfsast$ can be increased by as much as a factor of $40\%$ in comparison with the prediction for the \SE{} Cosmology.  
The larger $\kfsast$ will tend to weaken the lower bound on the mass of warm wave dark matter by a commensurate factor.  
If future observations at a sensitivity comparable to CMB-HD would detect free streaming suppression in the small-scale power spectrum, it could be possible to distinguish the \vEDE{} and \SE{} Cosmologies if the associated free streaming scales differ at the $10\%$ level.  
Uncertainties associated with theoretical modeling must be reduced to a commensurate level. 

%==================================
% Summary and conclusion
%==================================
\section{Summary and conclusion}
\label{sec:conclusion}

%=========
%Summary
In this work we have studied the impact of modified cosmic expansion histories on the free streaming of warm wave dark matter. 
The key elements of our work can be summarized as follows.  
\vspace{-0.5cm}
\begin{itemize}
%---
    \item  In \sref{sec:model_0} we revisit the observational constraints on warm wave dark matter in the standard expansion (\SE{}) cosmology, which has been studied in earlier work.  Using measurements of the linear matter power spectrum, which were inferred from observations of the \Lya{} forest spectra with eBOSS, we derive limits on the \WWDM{} parameter space ($\qast$, $m$), which appear in \fref{fig:model_0_params}.  Constraints arise from a combination of the white-noise effect and the free-streaming effect.  In the corner of parameter space where these two effects compete against one another, we find that the limit is slightly weaker than one would infer from the intersection of the two separate limits.  For $\qast \approx 300 \hMpc$ we find $m > 4 \times 10^{-21} \eV$, and for larger values of $\qast$ the limit is $\qast / m < 5 \times 10^{-8}$. We find that an experiment with the projected sensitivity of CMB-HD, would be sensitive to $m \approx 10^{-17} \eV$, and thereby provide a powerful probe of warm wave dark matter.  
%---
    \item  In \sref{sec:model_1} we study a modified cosmological expansion history that includes a period of early matter domination (\EMD{}) prior to the epoch of Big Bang Nucleosynthesis (\BBN{}).  Since a significant modification to the free streaming scale is only possible if the dominant modes are nonrelativistic during the \EMD{}, and since the \EMD{} must occur sufficiently early (before \BBN{}), we find that free streaming is only modified if the dark matter mass is $m > 10^{-15} \eV$, see \fref{fig:model_1}, but for such large mass the free streaming  wavenumber is too large to be observationally accessible.  
%---
    \item  In \sref{sec:model_2} we study a model of early dark energy (\EDE{}), motivated by the Hubble tension.  Since the new energy component enters at recombination, which is probed with high precision by \CMBR{} observations, earlier work has shown that the energy fraction in the \EDE{} can be no larger than $5\%$ of the total energy.  We find that this small energy enhancement translates to a sub-percent increase in the free streaming wavenumber $\kfsast$.  The detection of this subtle effect would require both extremely precise measurements and theoretical modeling. 
%---
    \item  Finally in \sref{sec:model_3} we study the recently-proposed scenario known as very early dark energy (\vEDE{}), which supposes that a new energy component dominates between the epochs of \BBN{} and recombination, and subsequently decays away without an associated injection of entropy to the Standard Model plasma.  We find that this modified cosmic history could significantly impact the free streaming of warm wave dark matter.  For instance, across the slice of parameter space illustrated in \fref{fig:model_3}, the free streaming  wavenumber $\kfsast$ can be enhanced by as much as $40\%$ over the \SE{} cosmology.  This encouraging result motivates a more systematic exploration of the broader joint \vEDE{}+\WWDM{} parameter space. 
\end{itemize}

%=========
%Conclusion
Dark matter inhomogeneities on small length scales offer a powerful probe of both dark matter's nature and the cosmological expansion history. 
We find that measurements of \Lya{} forest spectra are already placing strong constraints on the momentum and mass scale of warm wave dark matter under the assumption of a Standard Expansion Cosmology.  
Future observations at the level of sensitivity expected for CMB-HD have the potential to not only discover the suppression of small-scale power that is expected for warm wave dark matter due to the free-streaming effect, but also to measure this suppression with such high precision as to reveal deviations from \SE{} Cosmology.  

%======================================
\acknowledgments
%https://jcap.sissa.it/jcap/help/JCAP/TeXclass/DOCS/JCAP-author-manual.pdf
%https://www.nsf.gov/pubs/policydocs/pappguide/nsf16001/aag_6.jsp
We are grateful to Mustafa Amin, Kimberly Boddy, Sten Delos, Adrienne Erickcek, Wayne Hu, Siyang Ling, Rayne Liu, Tristan Smith, and Huangyu Xiao for illuminating discussions of free streaming and cosmological observables.  
We especially thank Sten Delos for helping us to develop the argument that appears in footnote~\ref{footnote5}.
Preliminary versions of this work were presented at the PITT PACC workshop ``Non-Standard Cosmological Epochs and Expansion Histories'' as well as the ``Theoretical Astroparticle and Cosmology Symposium'' at UT-Austin, and we are grateful to the workshop participants for feedback.  
This research was supported (in part: A.J.L.) by grant no. NSF PHY-2309135 to the Kavli Institute for Theoretical Physics (KITP).
This material is based upon work supported (in part: A.J.L. and M.V.) by the National Science Foundation under Grant No.~PHY-2412797.  

\bibliographystyle{JHEP}
\bibliography{refs}

\end{document}